\documentclass[aps,prd,twocolumn,superscriptaddress,nofootinbib]{revtex4}

\usepackage{graphicx}
\usepackage{amssymb, amsmath,mathtools,amsthm}
\newtheoremstyle{dotless}{}{}{\itshape}{}{\bfseries}{}{ }{}
\theoremstyle{dotless}
\makeatletter
\def\@endtheorem{\endtrivlist}
\makeatother
\newtheorem*{proposition*}{}

\usepackage[colorlinks=false,urlcolor=black]{hyperref}

\newcommand{\be}{\begin{equation}}
\newcommand{\bea}{\begin{align}}
\newcommand{\eea}{\end{align}}
\newcommand{\beq}{\begin{equation}}
\newcommand{\ee}{\end{equation}}
\newcommand{\eeq}{\end{equation}}

\newcommand{\poincare}{Poincar\'{e}}

\def\hthree{${\mathbb{H}}^3$}
\def\ethree{${\mathbb{E}}^3$}

\begin{document}


\title{Strong-field tidal distortions of rotating black holes:\\
  III. Embeddings in hyperbolic 3-space}



\author{Robert F. Penna}
\email[]{rp2835@columbia.edu}
\affiliation{Center for Theoretical Physics, Columbia University,
New York, New York 10027, USA}
\author{Scott A. Hughes}
\author{Stephen O'Sullivan}
\affiliation{Department of Physics and Kavli Institute for Astrophysics and Space Research,
Massachusetts Institute of Technology, Cambridge, Massachusetts 02139, USA}


\date{\today}

\begin{abstract}
In previous work, we developed tools for quantifying the tidal distortion of a black hole's event horizon due to an orbiting companion.  These tools use techniques which require large mass ratios (companion mass $\mu$ much smaller than black hole mass $M$), but can be used for arbitrary bound orbits, and for any black hole spin.  We also showed how to visualize these distorted black holes by embedding their horizons in a global Euclidean 3-space, \ethree.  Such visualizations illustrate interesting and important information about horizon dynamics.  Unfortunately, we could not visualize black holes with spin parameter $a_* > \sqrt{3}/2 \approx 0.866$: such holes cannot be globally embedded into \ethree.  In this paper, we overcome this difficulty by showing how to embed the horizons of tidally distorted Kerr black holes in a hyperbolic 3-space, \hthree.  We use black hole perturbation theory to compute the Gaussian curvatures of tidally distorted event horizons, from which we build a two-dimensional metric of their distorted horizons.  We develop a numerical method for embedding the tidally distorted horizons in \hthree.  As an application, we give a sequence of embeddings into \hthree\ of a tidally interacting black hole with spin $a_*=0.9999$.  A small amplitude, high frequency oscillation seen in previous work shows up particularly clearly in these embeddings.

\end{abstract}


\maketitle

\section{Introduction}
\label{sec:intro}


A body orbiting a black hole raises tidal bulges on the event horizon,
just as the moon raises ocean tides on the Earth.  Gravitational
torques due to these bulges cause the black hole and orbiting body to
exchange energy.  The direction of the energy exchange depends on the
relative rotation rate of the horizon and the orbit
\cite{Teukolsky:1974yv}.  For fluid bodies in Newtonian gravity, this
direction of energy exchange can be described in simple geometric
terms: If the angular velocity of the spinning body is faster than
that of the orbit ($\Omega_{\rm spin} > \Omega_{\rm orb}$), the bulges
lead the orbit's position, and energy flows from the body to the
orbit.  If the angular velocity of the body is slower than the orbit
($\Omega_{\rm spin} < \Omega_{\rm orb}$), the bulges lag the orbit,
and energy flows from the orbit to the body.

We cannot make such a simple connection between bulges and orbit
position for the tidal coupling of an orbit to a black hole's event
horizon.  In the limit of slow rotation and slow orbital velocity
\cite{Hartle:1974gy}, the Newtonian picture applies provided that we
reverse ``lag'' and ``lead'': the bulge leads the orbit for an orbit
that is slower than the hole's rotation, but lags when the orbit is
faster.  This counterintuitive exchange of ``lead'' and ``lag'' arises
from the horizon's teleological nature.  Determining whether an event
is inside or outside a horizon depends on that event's future history.
As such, an event horizon arranges its geometry in anticipation of the
stresses that it will feel in the future
\cite{Hartle:1974gy,Fang:2005qq,Damour:2009va,Poisson:2009di,Poisson:2009qj,Vega:2011ue,O'Sullivan:2014cba,Poisson:2014gka,O'Sullivan:2015lni}.

Various tools have been developed to understand the geometry of a
distorted black hole's event horizon.  A particularly useful tool has
been to develop {\it embeddings} of the distorted horizon.  An
embedding is a surface $r(\theta,\phi)$ drawn in some 3-dimensional
space whose geometry duplicates the geometry of the distorted black
hole's event horizon.  In the two previous papers in this series
\cite{O'Sullivan:2014cba,O'Sullivan:2015lni}, we developed tools to
embed the horizons of Kerr black holes in Euclidean 3-space, \ethree.
Unfortunately, these tools cannot be used if the black hole's
dimensionless spin parameter is $a_*>\sqrt{3}/2\approx 0.866$.  Such
holes have no global isometric embedding into
\ethree\ \cite{Smarr:1973zz}, so it is impossible to visualize their
tidal interactions in this way.

The goal of this paper is to address this issue by developing
techniques to isometrically embed tidally distorted black hole event
horizons into hyperbolic 3-space, \hthree.  All closed surfaces have
global isometric embeddings into \hthree\ (for some sufficiently small
choice of hyperbolic length scale), so in some sense it is a more
natural arena for visualizing surfaces than
\ethree\ \cite{pogorelov1964some}.  Embeddings of undistorted black
hole event horizons into \hthree\ have been considered by Gibbons et
al. \cite{Gibbons:2009qe}.  They used the \poincare\ half-plane model
for hyperbolic space.  We will use the \poincare\ ball model, which
preserves the symmetry of the undistorted horizon under reflection
across the equatorial plane.

Our motivation for this analysis is in part simply to complete the story developed in this paper's predecessors, Refs.\ \cite{O'Sullivan:2014cba} and \cite{O'Sullivan:2015lni}.  Although we developed techniques to quantify the tidal distortion of any Kerr black hole, our previous techniques only allowed us to visualize the distortion for those with spins $a_* \le \sqrt{3}/2$, a somewhat frustrating limitation.  Another motivation is to advocate on behalf of these hyperbolic spaces for visualizing the geometry of rapidly rotating black holes.  These spaces and the techniques we develop here may be useful for interpreting numerical relativity simulations involving rapidly rotating black holes.  Our results can be generalized to other surfaces in black hole spacetimes, such as apparent horizons and ergospheres.  In this context, we note that there is now solid observational evidence that black holes with $a_*>\sqrt{3}/2$ exist \cite{McClintock:2011zq}.  The ability to visualize can be a powerful aid to intuition.

Our analysis proceeds in three steps.  First, we use black hole perturbation theory to compute the Gaussian curvatures of tidally distorted black hole event horizons.  This machinery was developed in \cite{O'Sullivan:2014cba,O'Sullivan:2015lni}.  We use the Teukolsky equation \cite{Teukolsky:1973ha} to compute the perturbation to the Weyl curvature arising from a body of mass $\mu$ orbiting a black hole of mass $M$.  Our analysis assumes a large mass ratio, $\mu \ll M$, and that the orbit is bound, so that it can be completely described in the frequency domain.  From the Weyl curvature perturbation, it is a simple matter to compute the Gaussian curvature of the perturbed black hole's horizon \cite{O'Sullivan:2014cba}.

The second step of our procedure is to reconstruct the two-dimensional metric of the horizon from its Gaussian curvature.  We explain this step in Sec. \ref{sec:reconstruct}.  We use the fact that every metric on the 2-sphere has the form $g=e^{2u}\zeta^* g_0$, where $g_0$ is the unit round metric, $u$ is a conformal factor, and $\zeta$ is a diffeomorphism.  The Cartan equations of structure give a differential equation for $u$ which depends on the Gaussian curvature of the distorted horizon.  We linearize this equation in the system's mass ratio and solve for $u$ by expansion in spherical harmonics.

In the third step, we develop a numerical method for isometrically embedding the tidally distorted horizon, $(S^2,g)$, into \hthree.  We begin with an embedding of an undistorted horizon.  We triangulate the undistorted horizon and iteratively adjust the positions of the vertices until the induced metric matches the desired metric.  Our algorithm is based on a earlier method for embedding surfaces into \ethree\ \cite{nollert1996visualization,Ray:2015aia}. 

As an application, we show the tidal evolution of a black hole event horizon with spin parameter $a_*=0.9999$ using a series of embeddings into \hthree.  As has been seen in much previous work \cite{Hartle:1974gy,Fang:2005qq,Damour:2009va,Poisson:2009di,Poisson:2009qj,Vega:2011ue,O'Sullivan:2014cba,Poisson:2014gka,O'Sullivan:2015lni}, tidal bulges appear in anticipation of tidal forces, \emph{before} the forces reach the horizon.  This is a teleological effect reflecting the global nature of event horizons.  We also examine a fairly small amplitude but high frequency oscillation in the horizon's embedding which had been noted in a previous analysis of the horizon's curvature \cite{O'Sullivan:2015lni}, but whose origin remains somewhat mysterious.  As of yet, we have not clarified the physical origin of this oscillation, but highlight that the embedding illustrates its nature even more clearly.

The remainder of this paper is organized as follows.  We begin in Sec.\ \ref{sec:unpert} by reviewing unperturbed black holes and the basics of hyperbolic geometry: horospheres, geodesics, spheres, and equidistant curves in the \poincare\ ball.  We then explain in Section \ref{sec:reconstruct} how to reconstruct the metric of a perturbed black hole event horizon from its Gaussian curvature, $\kappa(\theta,\phi)$.  Section \ref{sec:embed} describes our numerical method for embedding distorted event horizons into hyperbolic space; Sec. \ref{sec:example} gives an application of these techniques to a tidally interacting black hole with $a_*=0.9999$. Supporting calculations are collected in Appendices \ref{sec:embedapp}--\ref{sec:edge}.

\section{Kerr black holes and hyperbolic space}
\label{sec:unpert}

Spinning, asymptotically flat black holes are described by the Kerr metric \cite{Kerr:1963ud}.  In Boyer-Lindquist coordinates \cite{Boyer:1966qh}, the event horizon is a null surface at radius $r_+ = M + \sqrt{M^2 - a^2}$, where $M$ and $a \equiv a_*M$ are the mass and spin parameters of the hole.  (We use units with $G = 1 = c$.)  The Kerr metric describes a black hole if $0 \le a_* \le 1$, and reduces to the Schwarzschild metric for $a_* = 0$.

Constant time slices of the event horizon are two-spheres, $S^2$.  The geometry of the horizon is slicing-independent because it is a stationary null surface.  The two-dimensional metric of the horizon on constant-time slices is \cite{Smarr:1973zz}
\beq\label{eq:gbar} \overline{ds}^2 = \eta^2
          [f^{-1}(\mu)d\mu^2+f(\mu)d\phi^2], \eeq where
\begin{align}
f(\mu) &= \frac{1-\mu^2}{1-\beta^2 (1-\mu^2)},\\
\eta &= \sqrt{r_+^2+a^2},\\
\beta &= \frac{a}{\sqrt{r_+^2 + a^2}}.
\end{align}
The quantity $\mu$ is related to the Boyer-Lindquist polar angle $\theta$ by $\mu \equiv \cos\theta$ (so $-1\leq \mu \leq 1$), and $\phi$ is the Boyer-Lindquist axial angle ($0 \leq \phi \leq 2\pi$).

The Gaussian curvature of the horizon is \cite{Smarr:1973zz}
\beq
\bar{\kappa}=-\frac{1}{2}f''(\mu)/\eta^2 = \frac{1-\beta^2(1+3\mu^2)}{\eta^2[1-\beta^2(1-\mu^2)]^3}.
\eeq
When $a=0$, the Gaussian curvature is $\bar{\kappa}=1/r_+^2$ and the horizon is a 2-sphere.  As the spin increases, the horizon becomes flattened, just as Newtonian fluid bodies are flattened by centrifugal forces.  The curvature increases at the equator and decreases at the poles.  At $a_*=\sqrt{3}/2\approx 0.866$, the horizon is flat ($\bar\kappa = 0$) at the poles $(\mu = \pm1$).  The polar regions become negatively curved ($\bar\kappa < 0$) for $a_*>\sqrt{3}/2$.

When the poles are negatively curved, the horizon has no global isometric embedding into Euclidean 3-space, \ethree.  In \ethree, the Gaussian curvature at a point on a surface is
\beq
\kappa = \kappa_1 \kappa_2,
\eeq
where $\kappa_1$ and $\kappa_2$ are the principal curvatures.  Negative curvature requires $\kappa_1$ and $\kappa_2$ to have opposite signs, so that negatively curved regions are locally saddle shaped.  However, rotational symmetry about the axis through the poles requires $\kappa_1 = \kappa_2$ at these points.  These conditions are clearly inconsistent, so we cannot surfaces with negatively curved poles in \ethree.

\begin{figure*}
\includegraphics[width=0.3\textwidth]{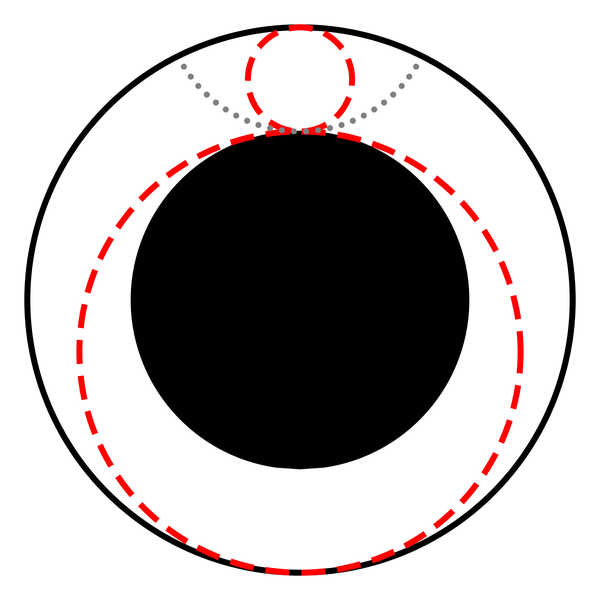}
\includegraphics[width=0.3\textwidth]{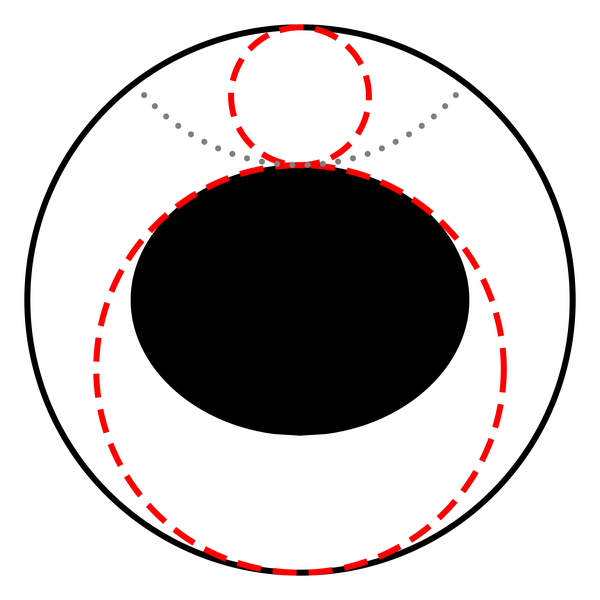}
\includegraphics[width=0.3\textwidth]{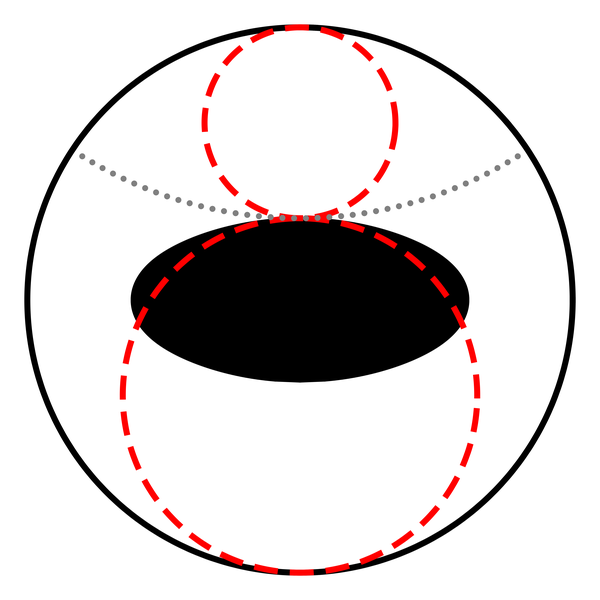}
\caption{Embeddings of undistorted black holes with spins  $a_*=0$ (left), $\sqrt{3}/2$ (center), and $1$ (right) into the \poincare\ ball.  We set $M=\ell=1$.  Horocycles (dashed red circles), geodesics (dotted gray curves), and the hyperbolic boundary at $r=1$ (black circles) are indicated.  The $a_*=1$ black hole is negatively curved at the north and south poles and has no global isometric embedding into Euclidean 3-space.}
\label{fig:bhs}
\end{figure*}

In \hthree, the (intrinsic) Gaussian curvature at a point on a surface is \cite{spivak1981comprehensive}
\beq
\kappa = \kappa_1 \kappa_2 - 1/\ell^2,
\eeq
where $\ell$ is the hyperbolic length scale (defined below).  The new term on the right-hand side contributes negatively to the Gaussian curvature, so negative Gaussian curvature is possible even if the principal curvatures $\kappa_1$ and $\kappa_2$ have the same sign.  In particular, in \hthree\ we can make surfaces of revolution with negative Gaussian curvature at the poles.  Thus, global isometric embeddings of rapidly spinning black hole horizons can be constructed in \hthree\ \cite{Gibbons:2009qe}.
The metric of \hthree\ in \poincare\ ball coordinates is
\beq\label{eq:ball}
ds_{{\mathbb{H}}^3}^2 = \frac{4\ell^2}{(1-r^2)^2}(dr^2+r^2 d\theta^2+r^2 \sin^2\theta d\phi^2),
\eeq
where $\ell$ is the hyperbolic length scale and $0\leq r< 1$.  The boundary at $r=1$ is an infinite proper distance from all points in \hthree.  It is straightforward to embed surfaces of revolution such as the horizon \eqref{eq:gbar} into the \poincare\ ball \eqref{eq:ball} (see Appendix \ref{sec:embedapp}).  It is possible to find global isometric embeddings for all black hole spins by making $\ell$ sufficiently small.  Figure \ref{fig:bhs} shows the results.  We set $M=\ell=1$ and consider embeddings with $a_*=0, \sqrt{3}/2, 1$.  The boundary of \hthree\ at $r=1$ is indicated by a black circle.  As the spin parameter $a_*$ increases, the curvature of the black hole horizon increases at the equator and decreases at the poles.

To understand these visualizations, it is helpful to compare the horizons with horospheres, geodesics, spheres, and equidistants in \hthree.  The dashed red circles in Figure \ref{fig:bhs} are ``horocycles,'' circles with one point on the boundary of \hthree.  Surfaces of revolution generated by horocycles are horospheres.  Horospheres have zero Gaussian curvature.  So one can intuit the sign of the curvature at the north and south poles of the black hole by making comparisons  with horospheres (see Figure \ref{fig:bhs}).  If the horizon is more curved than the tangent horosphere then it is positively curved, and if it is less curved than the tangent horosphere it is negatively curved.  The $a_*=0$ black hole in Figure \ref{fig:bhs} is positively curved at the poles, the $a_*=\sqrt{3}/2$ black hole is flat at the poles, and the $a_*=1$ black hole is negatively curved at the poles.

The dotted gray arcs in Figure \ref{fig:bhs} are geodesics.  In the \poincare\ ball, geodesics are arcs of circles which are perpendicular to the boundary.  Surfaces of revolution generated by geodesics have Gaussian curvature $\kappa = -1/\ell^2$.  This is the smallest possible curvature at the poles of a surface of revolution in \hthree\ (for fixed $\ell$).  

A sphere in \hthree\ is a collection of points with the same proper distance from a given point.  Hyperbolic and Euclidean spheres have the same images but they have different centers. The hyperbolic center is closer to the boundary at $r=1$.  Consider a sphere centered on the origin of the \poincare\ ball with coordinate radius $r$.  Its proper radius is
\beq\label{eq:rcirc}
R = \frac{2}{\ell} {\rm \thinspace arctanh\thinspace} r,
\eeq
and its Gaussian curvature is
\beq\label{eq:kcirc}
\kappa = \frac{1}{\ell^2} (\coth^2(R/\ell)-1).
\eeq
For $R\ll \ell$, the Gaussian curvature is $\kappa \approx 1/R^2$, just as in Euclidean space.  In the limit $R\rightarrow \infty$, the Gaussian curvature $\kappa\rightarrow 0$. 
The $a_*=0$ black hole in Figure \ref{fig:bhs} is a round sphere.  In \poincare\ coordinates, it has radius $r=(\sqrt{5}-1)/2\approx 0.618$.  (Amusingly, this is the (inverse) golden ratio).  Plugging $r$  into \eqref{eq:rcirc}--\eqref{eq:kcirc} gives $\kappa=1/r_+^2$, as expected.    

Equidistants from the $z$-axis of the \poincare\ ball are arcs of circles which intersect the boundary at $z=-1$ and $z=+1$ (Cartesian coordinates $(x,y,z)$ are related to spherical coordinates $(r,\theta,\phi)$ in the usual way).  All spheres which are centered on the $z$-axis and touch a given equidistant at a single point have the same proper radius and Gaussian curvature.  This means objects with the same proper size appear smaller near the boundary of \hthree. One can use equidistants to move surfaces around \hthree\ while preserving their proper sizes and curvatures.  This is useful for comparing surfaces at different locations.

\begin{figure}
\includegraphics[width=0.48\columnwidth]{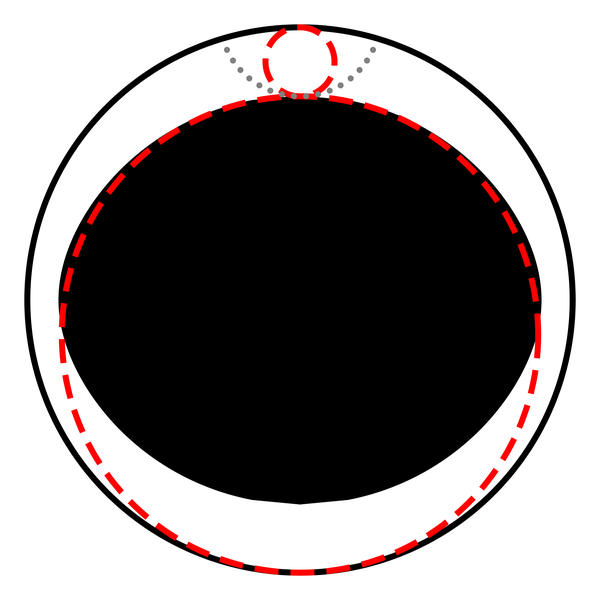}
\includegraphics[width=0.48\columnwidth]{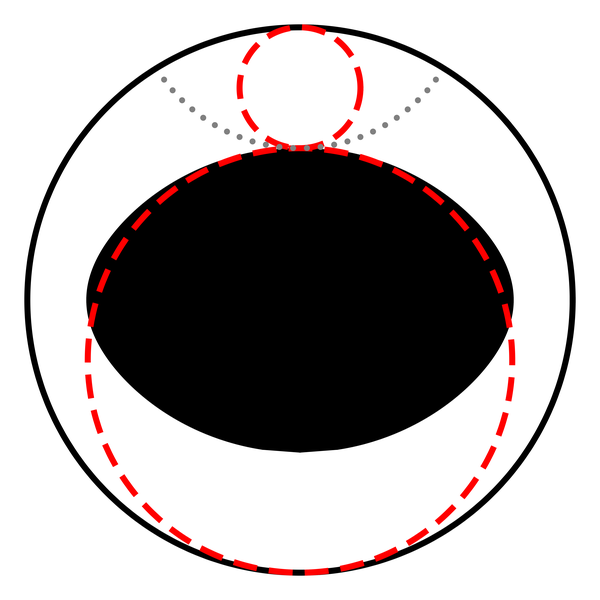}\\
\includegraphics[width=0.48\columnwidth]{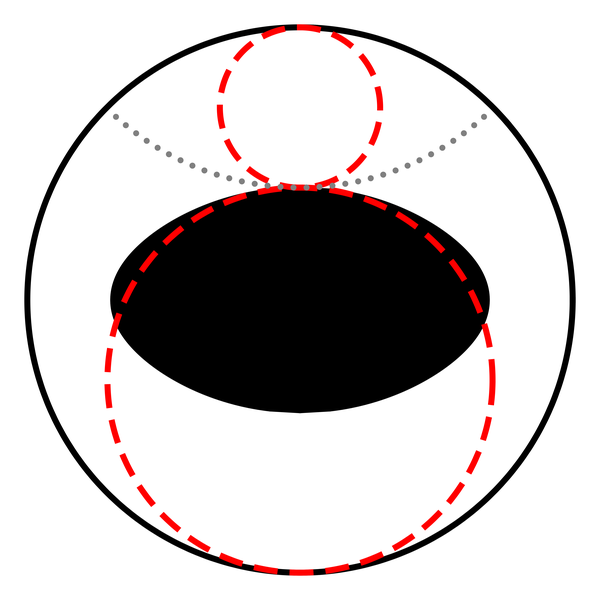}
\includegraphics[width=0.48\columnwidth]{a1.png}
\caption{Embeddings of the $a_*=M=1$ black hole into \poincare\ balls with $\ell=1/4$ (top left), $1/2$ (top right), $3/4$ (bottom left), and $1$ (bottom right).}
\label{fig:bh1}
\end{figure}

The most negatively curved regions in Figure \ref{fig:bhs} are the north and south poles of the $a_*=1$ black hole, where $\bar{\kappa}=-1/2$.    In Figure \ref{fig:bhs}, we set $\ell=1$. However, this surface can be embedded into \hthree\ for any $\ell \leq 1/\sqrt{-\bar{\kappa}}= \sqrt{2}$. Figure \ref{fig:bh1} shows embeddings of the same surface for different $\ell$.  The apparent shape of the black hole depends on $\ell$.    Only the relationship between the apparent shape and the hyperbolic boundary at $r=1$ is meaningful.

\begin{figure}
\includegraphics[width=\columnwidth]{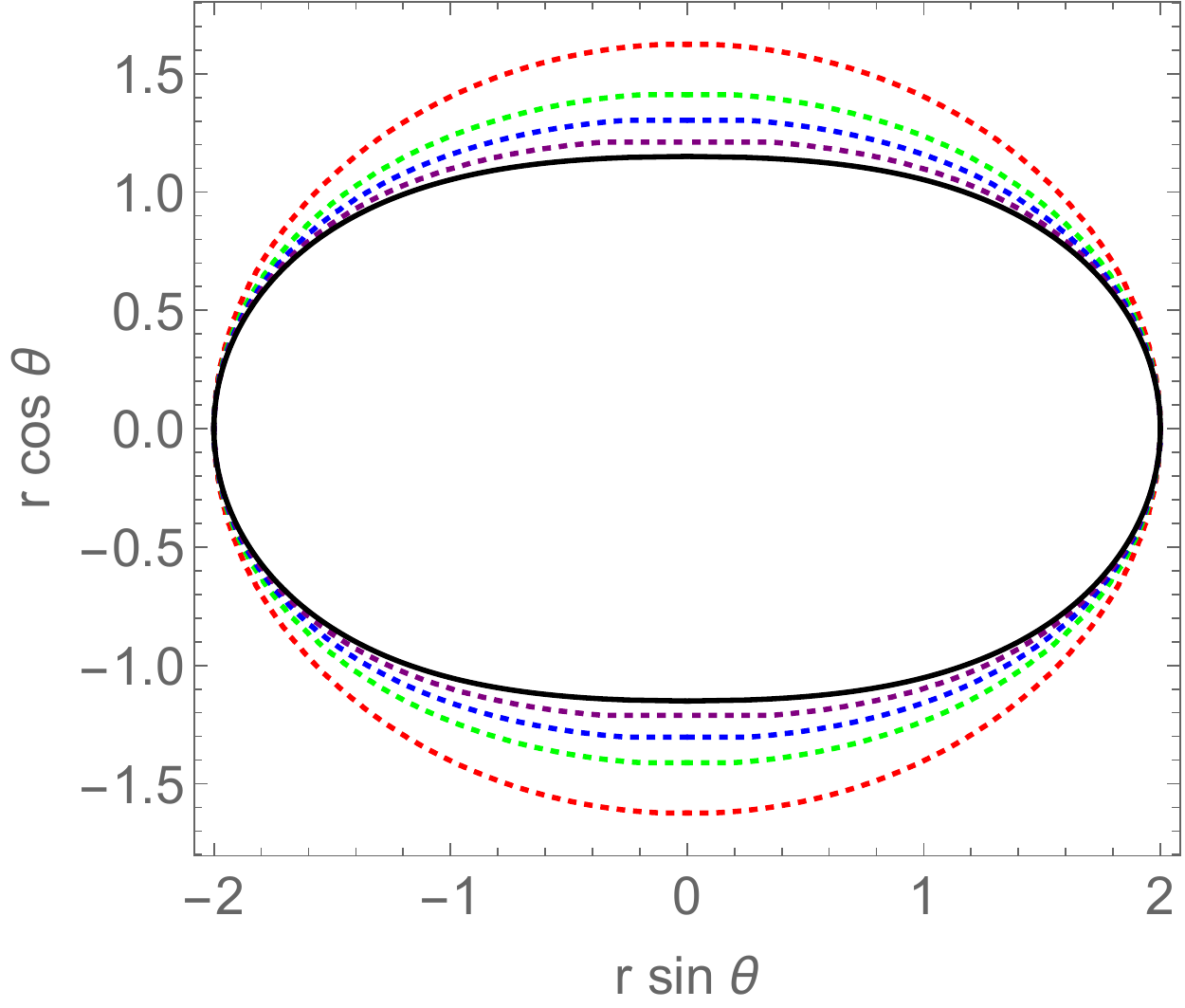}\\
\hspace{0.4 in}
\includegraphics[width=\columnwidth]{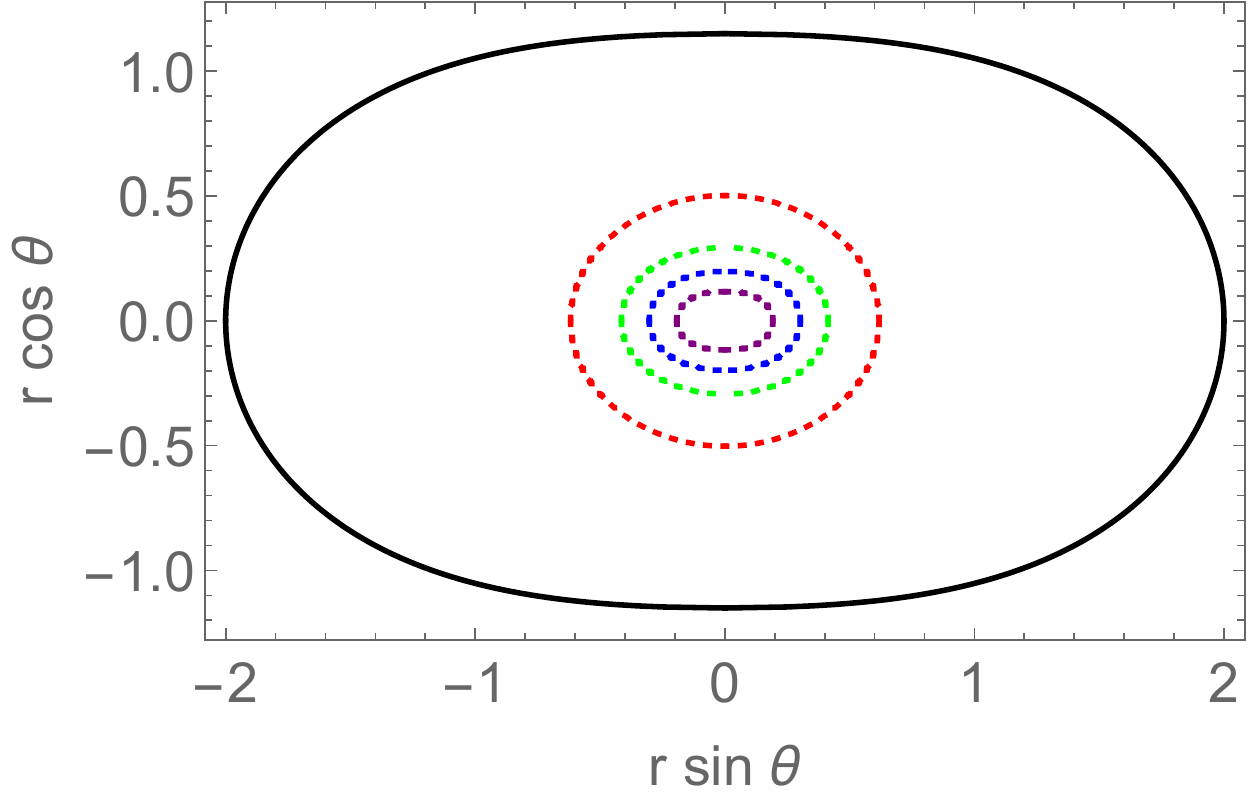}
\caption{Embeddings of an unperturbed $a_*=0.85$ black hole into Euclidean 3-space (solid black curve) and hyperbolic 3-space (dotted curves) for $\ell=1$ (red), $2$ (green), $3$ (blue), and $5$ (purple).  In the top panel, the hyperbolic radial coordinate, $r$, has been rescaled and the hyperbolic boundary is at $r\approx 3.2$ (red), $4.8$ (green), $6.6$ (blue), and $10.4$ (purple).  In the bottom panel, the hyperbolic boundary is fixed at $r=1$.}
\label{fig:compare}
\end{figure}

To understand the relationship between hyperbolic and Euclidean embeddings, it is helpful to compare embeddings of the same horizon into both spaces.  Both embeddings exist when $a_*\leq \sqrt{3}/2$.  Figure \ref{fig:compare} shows a black hole horizon with $a_*=0.85$ embedded in Euclidean 3-space and hyperbolic 3-space, for a sequence of hyperbolic length scales, $\ell$.   The hyperbolic embeddings converge to the Euclidean embedding in the limit $\ell\rightarrow \infty$.  Already at $\ell=5$ they are nearly the same.  Decreasing $\ell$ has the effect of ``puffing up'' the surface so that its apparent shape becomes increasingly round.  In the top panel of Figure \ref{fig:compare}, we have rescaled the hyperbolic radial coordinate so that all embeddings have the same coordinate size at the equator.  This moves the hyperbolic boundary to $r>1$ (see the caption of Figure \ref{fig:compare} for details).  The bottom panel of Figure \ref{fig:compare} shows the same series of embeddings without rescaling, so that the hyperbolic boundary is fixed at $r=1$.

\section{Perturbing the black hole and reconstructing the horizon metric}
\label{sec:reconstruct}

Consider a black hole perturbed by an orbiting body.  As described in detail in Ref.\ \cite{O'Sullivan:2014cba}, we use the Teukolsky equation to compute the Weyl curvature scalar $\psi_0$, from which we compute the Gaussian curvature of the perturbed horizon,
\beq\label{eq:kappabh}
\kappa (\theta,\phi) = \bar{\kappa}(\theta,\phi) + \epsilon \tilde{\kappa}(\theta,\phi)\;.
\eeq
In this equation, $\bar{\kappa}$ is the unperturbed curvature, $\epsilon$ is the mass ratio.  We assume $\epsilon \ll 1$ and $\tilde{\kappa}$ is independent of $\epsilon$.  Our goal in this section is to describe how to reconstruct the two-dimensional metric of the perturbed event horizon from $\kappa$.

Constant-time slices of the horizon have the topology $S^2$.  Every metric on $S^2$ has the form\footnote{This follows from the uniformization theorem for surfaces.}
\beq\label{eq:uniform}
g = e^{2u} \zeta^* g_0,
\eeq
where $g_0$ is the unit round metric, $u$ is a scalar function, $\zeta$ is a diffeomorphism, and  $\zeta^* g_0$ is the pull-back of $g_0$ by $\zeta$.  Our task is to find $u$ and $\zeta$ such that the Gaussian curvature of $g$ is $\kappa$.  

The black hole is weakly perturbed, so let
\beq\label{eq:upert}
u = \bar{u} + \epsilon \tilde{u},
\eeq
where $\bar{u}$ is the unperturbed conformal factor.  First we need to determine the unperturbed conformal factor $\bar{u}$ by comparing the unperturbed metric  \eqref{eq:gbar}
\beq\label{eq:gbar3}
\overline{ds}^2 = \eta^2 f(\mu)[f^{-2}(\mu)d\mu^2+d\phi^2],
\eeq
with the unit round metric:
\beq\label{eq:g0}
ds_0^2 = f_0(\mu')\left[f_0(\mu')^{-2}d\mu'^2+d\phi'^2\right],
\eeq
where $f_0(\mu')= 1-\mu'^2$.  The terms inside the square brackets are related by the coordinate change
\begin{align}
d\mu'/f_0(\mu')&=d\mu/f(\mu), \label{eq:barzeta1}\\ 
\phi' &= \phi.\label{eq:barzeta2}
\end{align}
This fixes the unperturbed diffeomorphism $\bar{\zeta}(\mu,\phi)=(\mu',\phi')$.
The terms outside the square brackets in eqs. \eqref{eq:gbar3}-\eqref{eq:g0}  fix $\bar{u}$:
\beq\label{eq:baru}
e^{2\bar{u}} = \frac{\eta^2 f}{f_0\circ \bar{\zeta}},
\eeq
where $f_0\circ\bar{\zeta}$ is the composition of $f_0$ with $\bar{\zeta}$.  Note that the numerator and denominator of the rhs are both functions of the unprimed coordinates.

Now we turn to the problem of computing the perturbed metric.  We define $\kappa'=\kappa\circ\zeta^{-1}$ and $u'=u\circ \zeta^{-1}$. These are just the functions $\kappa$ and $u$ evaluated in primed coordinates $(\mu',\phi')$.  They are related by
\beq\label{eq:kw}
\kappa' =  -e^{-2u'}(\Delta u' - 1),
\eeq
where $\Delta$ is the Laplacian with respect to the unit round metric (see Appendix \ref{sec:kw}).  

Equation \eqref{eq:kw} is a single differential equation for the two unknown functions, $\zeta$ and $\tilde{u}$.  For simplicity, we will seek solutions with $\zeta = \bar{\zeta}$.  This leaves one unknown, $\tilde{u}$, which appears nonlinearly in \eqref{eq:kw}.  We are only interested in weakly perturbed black holes, so we can use the fact that $\epsilon\ll 1$ to linearize \eqref{eq:kw} in the mass ratio.   Linearizing and rearranging gives
\beq\label{eq:kwlinear}
\Delta \tilde{u}' = -e^{2\bar{u}'}(2\tilde{u}'\bar{\kappa}'+\tilde{\kappa}'),
\eeq
which is a linear differential equation for the single unknown function $\tilde{u}'$.  We have set $\bar{u}'=\bar{u}\circ \bar{\zeta}^{-1}$ and $\bar{\kappa}'  =\bar{\kappa}\circ \bar{\zeta}^{-1}$, and we have used the relation
\beq\label{eq:kwbar}
\bar{\kappa}'=  -e^{-2\bar{u}'}(\Delta \bar{u}' - 1),
\eeq
which is proved in Appendix \ref{sec:kw}. 

Equation \eqref{eq:kwlinear} is solved by series expansion.  Let
\begin{align}
\tilde{u}'  &= \sum_{lm} a^m_l Y^m_l, \label{eq:alm}\\ 
-2e^{2\bar{u}'}\bar{\kappa}' &= \sum_{lm} b^m_l Y^m_l, \label{eq:blm}\\ 
-e^{2\bar{u}'}\tilde{\kappa}' &= \sum_{lm} c^m_l Y^m_l,\label{eq:clm}
\end{align}
where $Y^m_l$ are the usual spherical harmonics on the unit round sphere.  We choose the normalization $\int Y^m_l{Y^{m'}_{l'}}^*d\Omega = \delta_{ll'}\delta_{mm'}$.  Plugging into \eqref{eq:kwlinear} and integrating against ${Y^m_l}^*$ gives
\begin{align}\label{eq:linear}
-l(l+1)a^m_l &= \sum_{l_1l_2m_1m_2}a^{m_1}_{l_1}b^{m_2}_{l_2}\sqrt{\frac{(2l_1+1)(2l_2+1)}{4\pi(2l+1)}}\notag\\
&\cdot\langle l_1l_2;00|l_1l_2;l0\rangle\langle l_1l_2;m_1m_2|l_1l_2;lm\rangle\notag\\
&+c^m_l,
\end{align}
where the $\langle l_1l_2;m_1m_2|l_1l_2;lm\rangle$ are Clebsch-Gordan coefficients \cite{sakurai}.  Equation \eqref{eq:linear} is an infinite system of linear algebraic equations for the $a^m_l$.   We obtain a finite set of equations by truncating \eqref{eq:alm}-\eqref{eq:clm} at a finite $l=n$.  The finite set is solved numerically. 
This completes the solution for the perturbed metric on the horizon.  

To summarize, the perturbed metric is
\beq
g=e^{2(\bar{u}+\epsilon\tilde{u})}\bar{\zeta}^*g_0,
\eeq
where $g_0$ is defined by \eqref{eq:g0},  $\bar{\zeta}$ is defined by \eqref{eq:barzeta1}-\eqref{eq:barzeta2},  $\bar{u}$ is defined by \eqref{eq:baru}, and $\tilde{u}=\tilde{u}'\circ \bar{\zeta}$ is obtained by solving \eqref{eq:linear} for the $a^m_l$.  By construction, the Gaussian curvature of $g$ is Eq.\ \eqref{eq:kappabh}, as desired.

\subsection{Example: perturbed Schwarzschild}
\label{sec:schwarz}

The simplest example is a perturbed Schwarzschild black hole ($a_*=0$).  In this case,
$e^{-2\bar{u}} = 1/\eta^2= 1/(4M^2)=\bar{\kappa}$, and $\bar{\zeta}$ is the identity.  Equation \eqref{eq:clm} becomes
\beq
\tilde{\kappa}' = -\frac{1}{\eta^2}\sum_{lm}c^m_l Y^m_l.
\eeq
Let $\zeta=\bar{\zeta}={\rm id}$.  Then there is no distinction between primed and unprimed quantities, and the solution of \eqref{eq:linear} is simply
\beq
a^m_l = \frac{c^m_l}{2-l(l+1)}.
\eeq
This is singular when $\tilde{\kappa}$ involves first degree spherical harmonics.  In this case, the metric cannot be  in the same conformal class as the round metric ($\zeta \neq \bar{\zeta}$).  This is related to an obstruction first observed by \cite{kazdan74}.  However, the spin-two nature of the gravitational field guarantees that first degree spherical harmonics can always be eliminated from $\tilde{\kappa}$.

When $a_*\neq 0$, the obstructions to solving \eqref{eq:linear} are less clear.  In all the examples we have checked, we were able to find solutions for the perturbed horizon in the same conformal class as the unperturbed horizon (i.e., assuming $\zeta=\bar{\zeta}$).

\section{Embedding into \hthree}
\label{sec:embed}

In the previous section, we reconstructed the metric of the perturbed horizon, $g$, from its Gaussian curvature.  In this section, our goal is to isometrically embed $(S^2,g)$ into \hthree.  

We begin with an isometric embedding of the unperturbed horizon, $(S^2,\bar{g})$, obtained as in Appendix \ref{sec:embedapp}.  We triangulate this surface using a maximal planar graph.  The triangulation has $N$ vertices and $3N-6$ edges.  Our goal is to adjust the positions of the vertices so that the surface converges to $(S^2,g)$.   We use a modification of the method developed by \cite{nollert1996visualization} for embedding surfaces into Euclidean 3-space.

Consider an edge in the triangulation with endpoints $p$ and $q$.  Let the initial coordinates of $p$ and $q$ be 
\begin{align}
\vec{x}^0_p &= (x^0_p,y^0_p,z^0_p),\\
\vec{x}^0_q &= (x^0_q,y^0_q,z^0_q).
\end{align}  
Our goal is to find new positions,
\begin{align}
\vec{x}_p &= (x_p,y_p,z_p),\\
\vec{x}_q &= (x_q,y_q,z_q),
\end{align} 
such that the embedded surface approaches $(S^2,g)$. We then iterate the adjustment process until the surface converges to $(S^2,g)$.

Define the initial length of edge $pq$ to be 
\beq\label{eq:dist1}
l^0_{pq}=\frac{4\ell^2}{(1-r^2)^2}
\left[(x^0_p-x^0_q)^2-(y^0_p-y^0_q)^2-(z^0_p-z^0_q)^2\right],
\eeq
where $r$ is evaluated at the midpoint of the edge.  Equation \eqref{eq:dist1} defines the distance between $p$ and $q$ as the length along a line in \hthree.  A better (but more complicated) approach is to define distances using geodesics on the embedded surface. We present these improvements in Appendix \ref{sec:edge}.

The desired length of edge $pq$ is
\beq\label{eq:lpq}
l_{pq}  = e^{2\tilde{u}(\theta,\phi)}l^0_{pq},
\eeq
where $\theta$ and $\phi$ are evaluated at the midpoint of the edge.  Let
\begin{align}
\vec{x}_p &= \vec{x}^0_p + \delta \vec{x}_p,\\
\vec{x}_q &= \vec{x}^0_q + \delta \vec{x}_q.
\end{align}
Our goal is to choose $\delta \vec{x}_p$ and $\delta \vec{x}_q$ such that the new length of $pq$ is $l_{pq}$.

Linearizing \eqref{eq:lpq}  gives
\beq\label{eq:taylor}
l^0_{pq}  (e^{2\tilde{u}}-1) = \frac{\partial l^0_{pq}}{\partial \vec{x}^0_p} \cdot\delta \vec{x}_p
+\frac{\partial l^0_{pq}}{\partial \vec{x}^0_q} \cdot \delta \vec{x}_q.
\eeq
The derivatives on the rhs are computed using \eqref{eq:dist1}.  For example,
\beq
\frac{\partial l^0_{pq}}{\partial x^0_p} = 8 \ell^2\frac{ x^0_p-x^0_q}{(1-r^2)^2}
+ \frac{2x^0_pl^0_{pq}}{1-r^2}.
\eeq

Equation \eqref{eq:taylor} is a linear equation for the unknowns $\delta \vec{x}_p$ and $\delta \vec{x}_q$.  Each edge in the triangulation gives one such equation, so we have $3N-6$ equations.   The total number of unknowns is $3N$ (three coordinates,  $\delta \vec{x}_p$, for each vertex $p$ in the triangulation), so we need six more equations.  These are provided by the six components of the constraint equations
\begin{align}
\sum_p \delta \vec{x}_p &= 0,\label{eq:constraint1}\\
\sum_p \vec{x}^0_p \times \delta \vec{x}_p &=0,\label{eq:constraint2}
\end{align} 
which fix the overall position and orientation of the surface.   

Eqs. \eqref{eq:taylor} and \eqref{eq:constraint1}-\eqref{eq:constraint2} are a system of $3N$ linear equations for the $3N$ unknowns, $\delta x_p$.   We have solved these equations numerically.  Iterating gives a series of surfaces which converge toward $(S^2,g)$.

\section{Example}
\label{sec:example}

Consider an $a_*=0.9999$ black hole perturbed by an orbiting body, and let $\epsilon=1/125$ be the mass ratio.  We assume the perturbing body is on an equatorial orbit with semi-latus rectum $p/M=M$ and eccentricity $e=0.7$.  
  
We use the tools developed in \cite{O'Sullivan:2014cba,O'Sullivan:2015lni} to compute the Gaussian curvature of the perturbed horizon as a function of ``ingoing time'' $v$.  (``Ingoing time'' is a variant of the Boyer-Lindquist time coordinate that is well behaved on the event horizon.)  Then we solve \eqref{eq:linear} to reconstruct the metric of the perturbed horizon as a function of $v$.  We keep terms up to order $l=7$.   Finally, we iteratively solve \eqref{eq:taylor} and \eqref{eq:constraint1}-\eqref{eq:constraint2} for the embedding.  We use a triangulation with $40\times 40$ vertices, evenly spaced in $\cos\theta$ and $\phi$, and iterate the embedding equations 16 times.

\begin{figure*}
\includegraphics[width=0.32\textwidth]{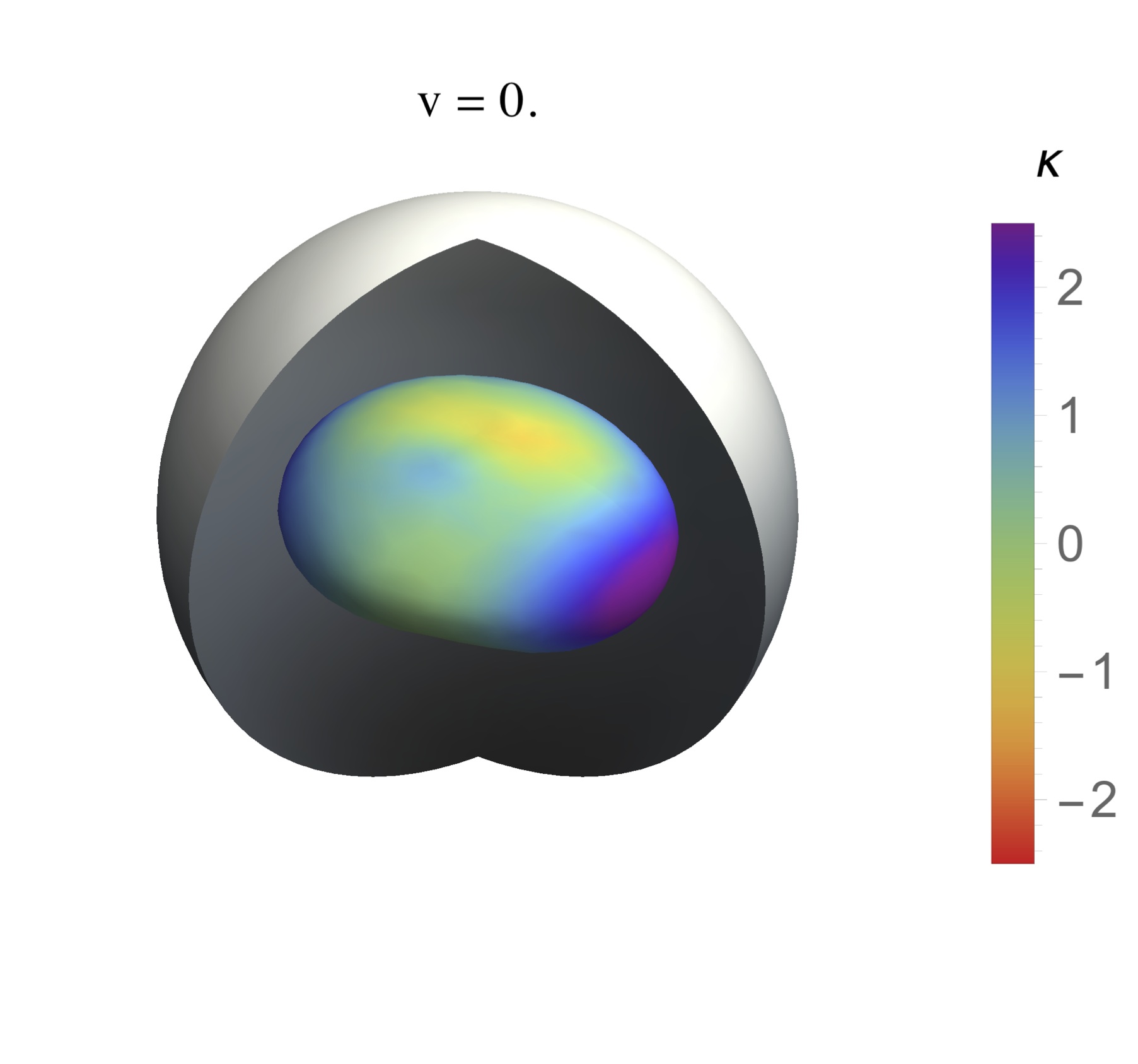}
\includegraphics[width=0.32\textwidth]{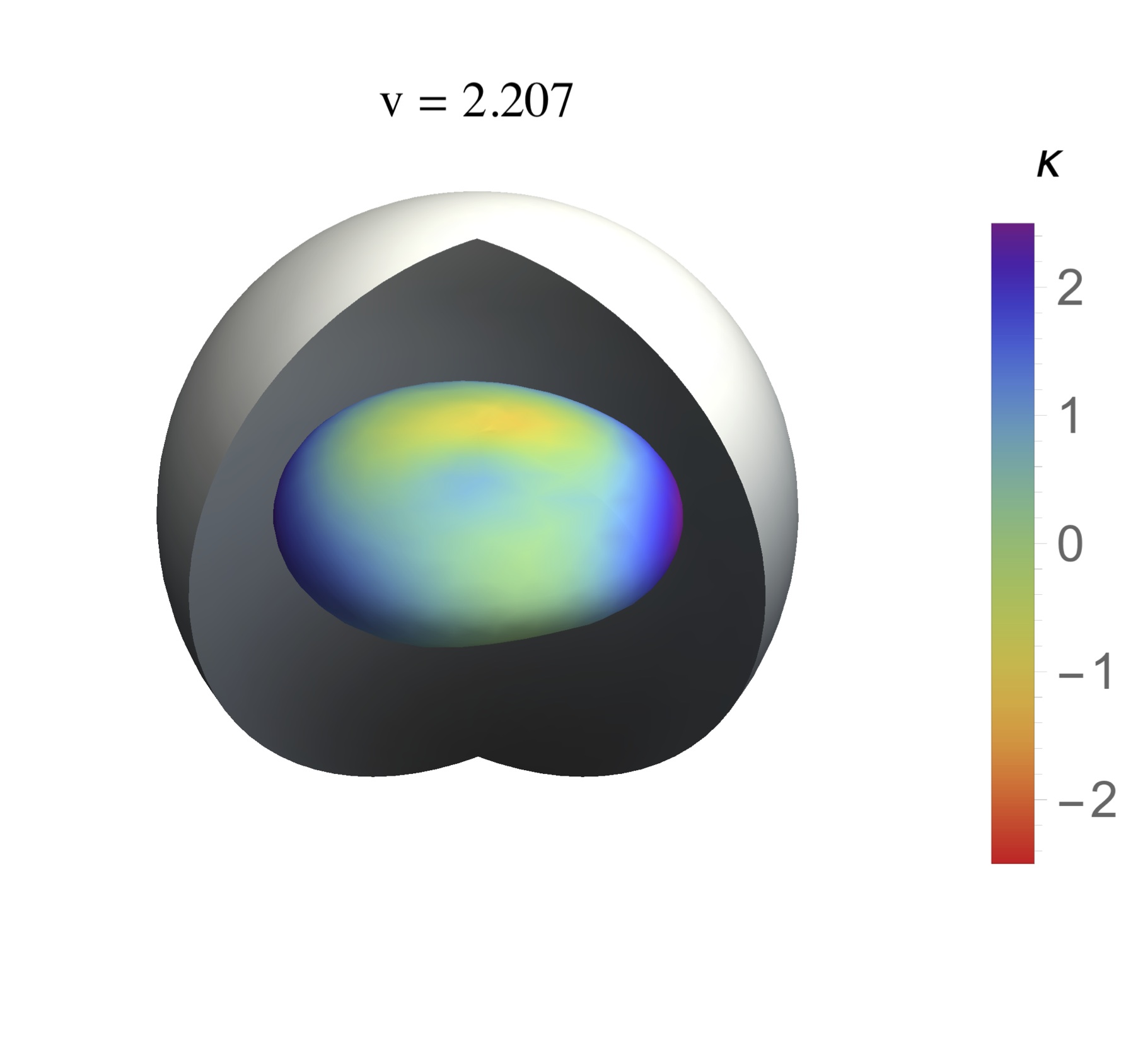}
\includegraphics[width=0.32\textwidth]{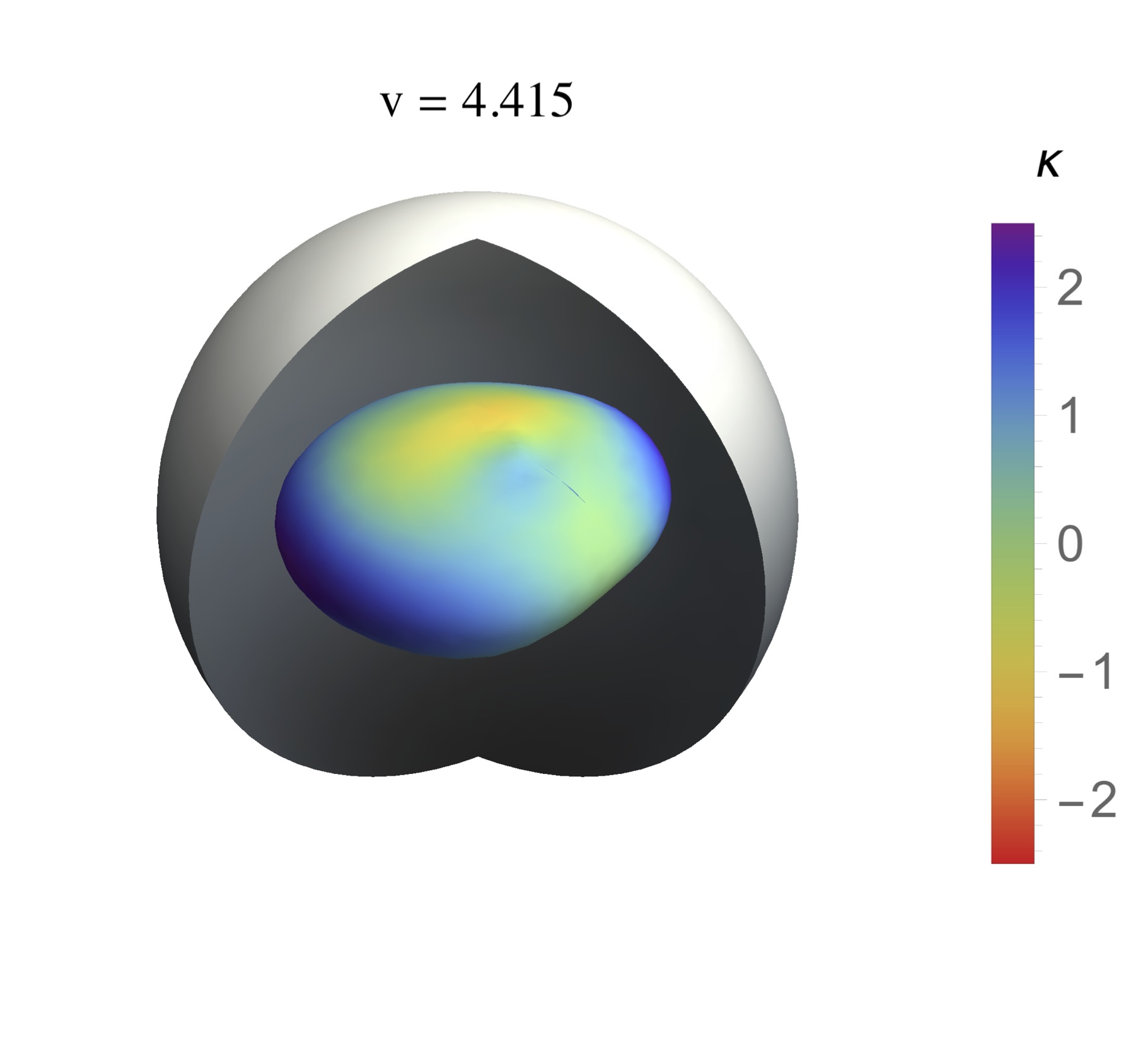}\\
\includegraphics[width=0.32\textwidth]{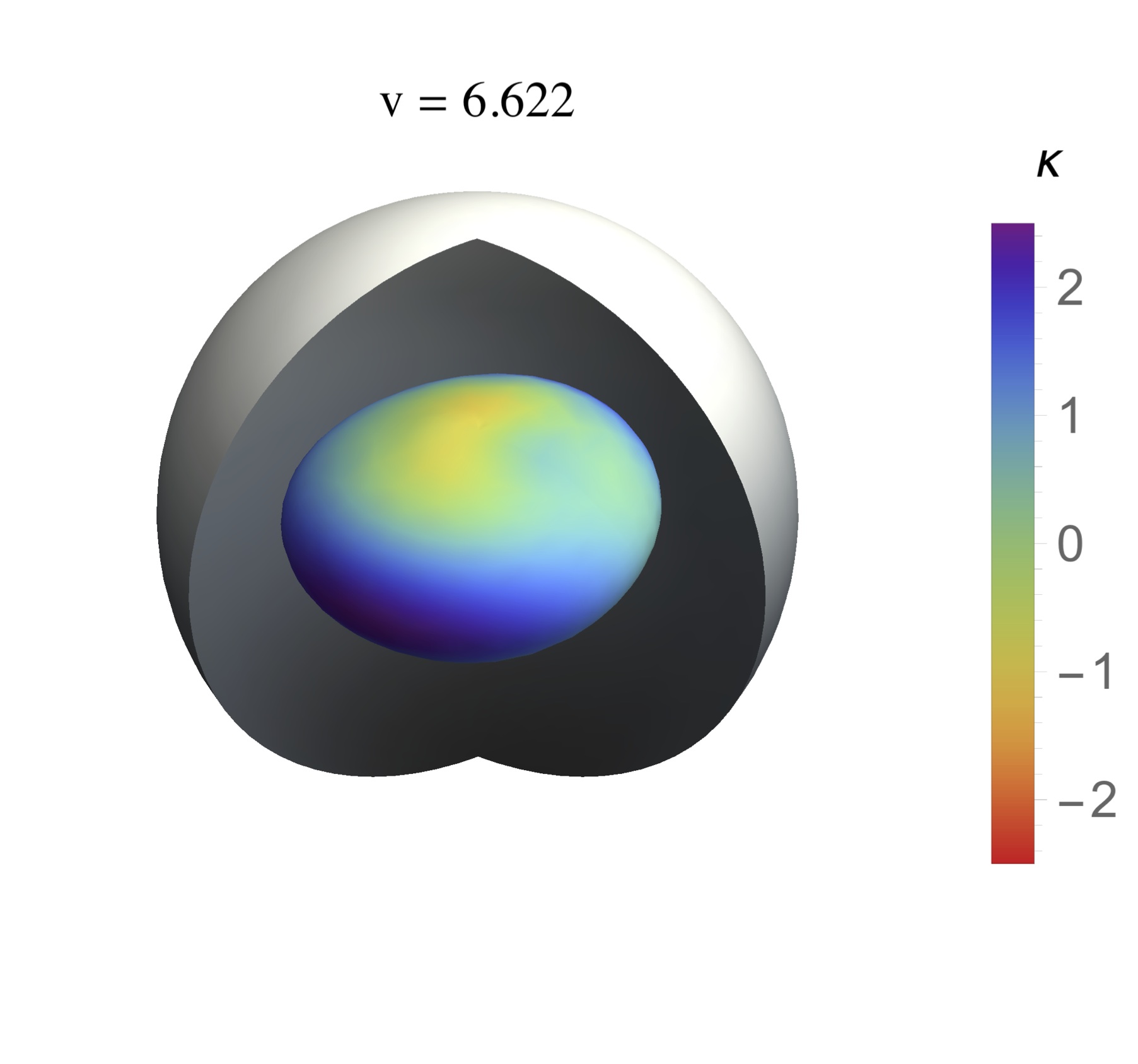}
\includegraphics[width=0.32\textwidth]{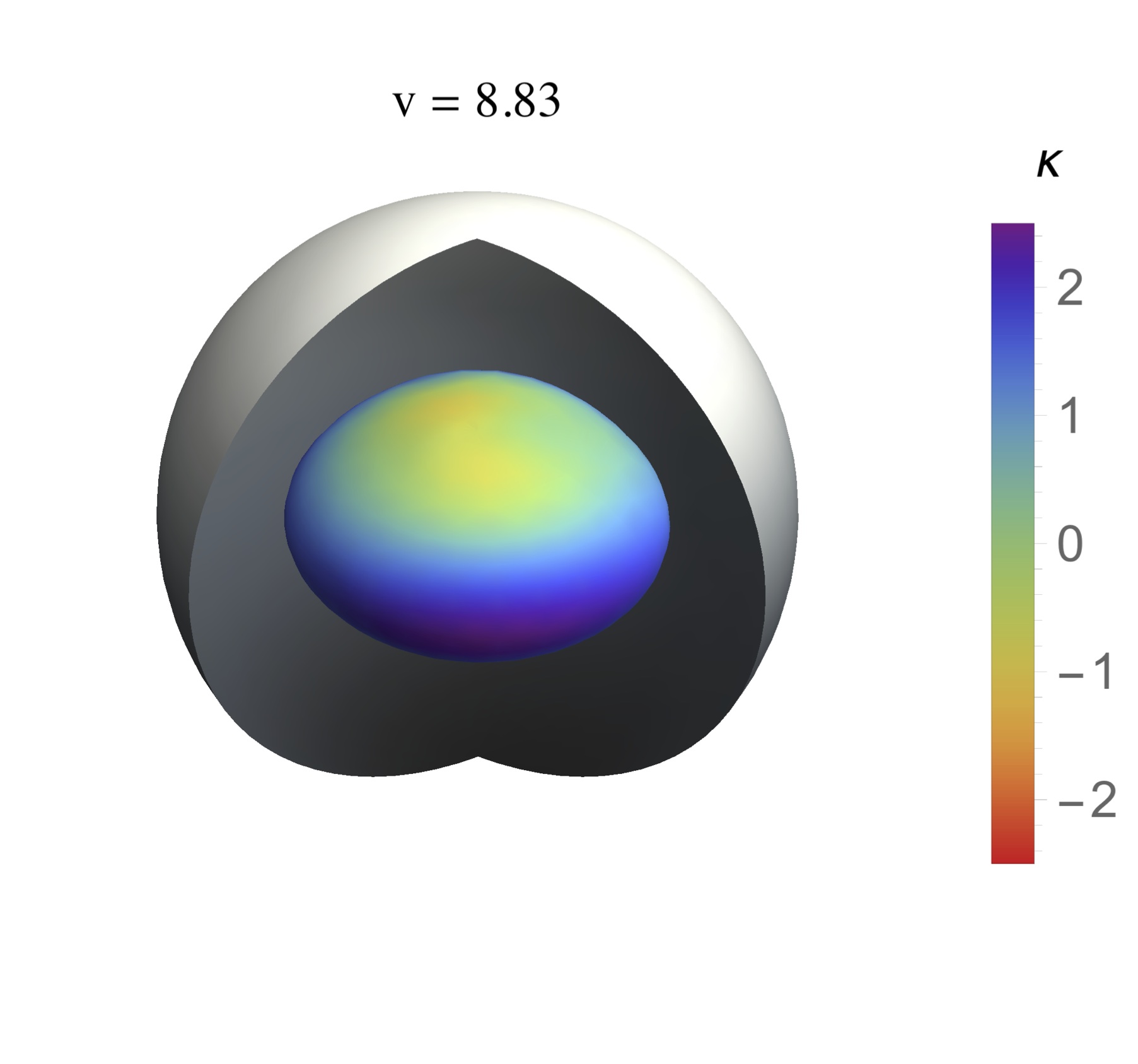}
\includegraphics[width=0.32\textwidth]{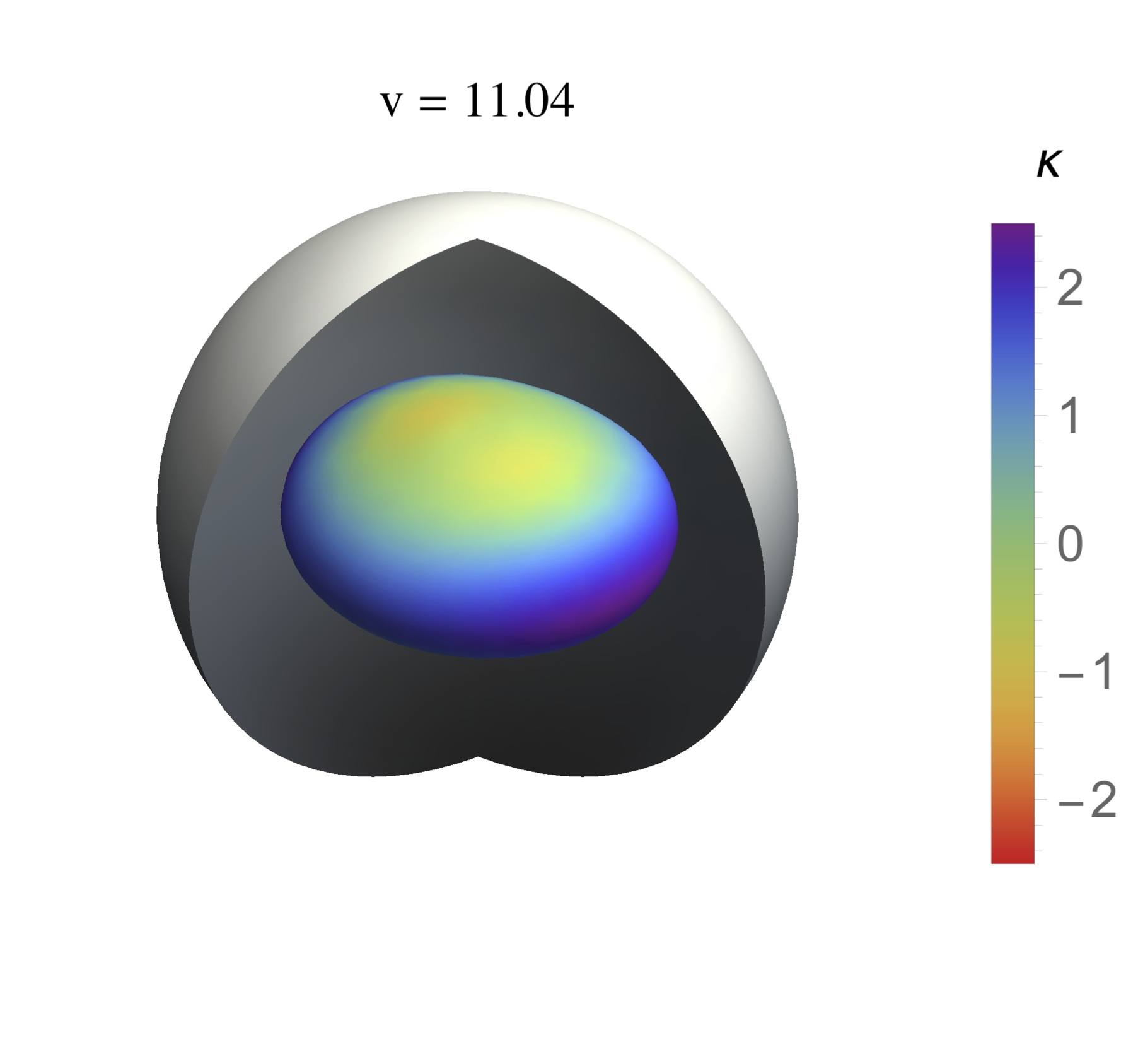}\\
\includegraphics[width=0.32\textwidth]{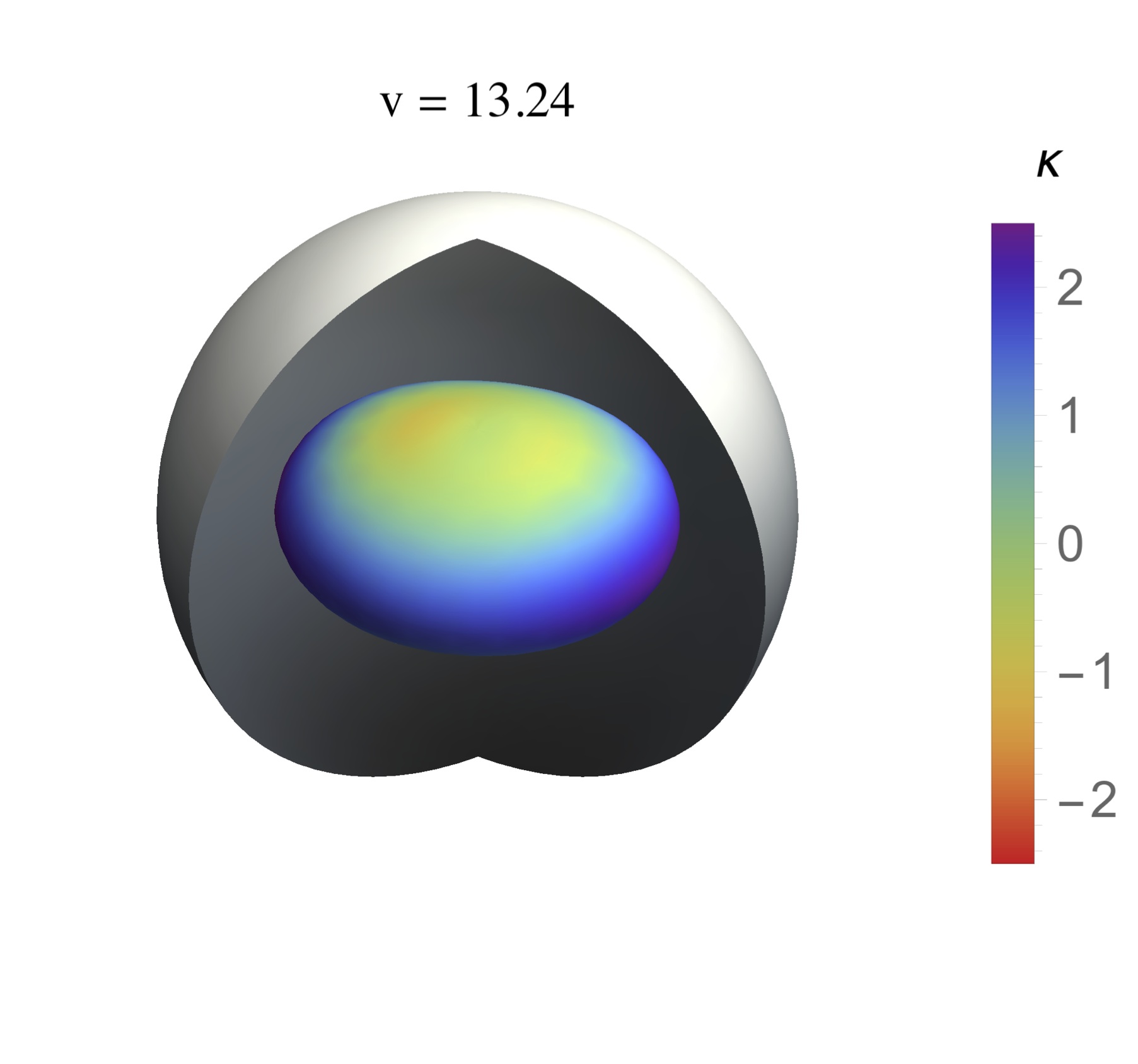}
\includegraphics[width=0.32\textwidth]{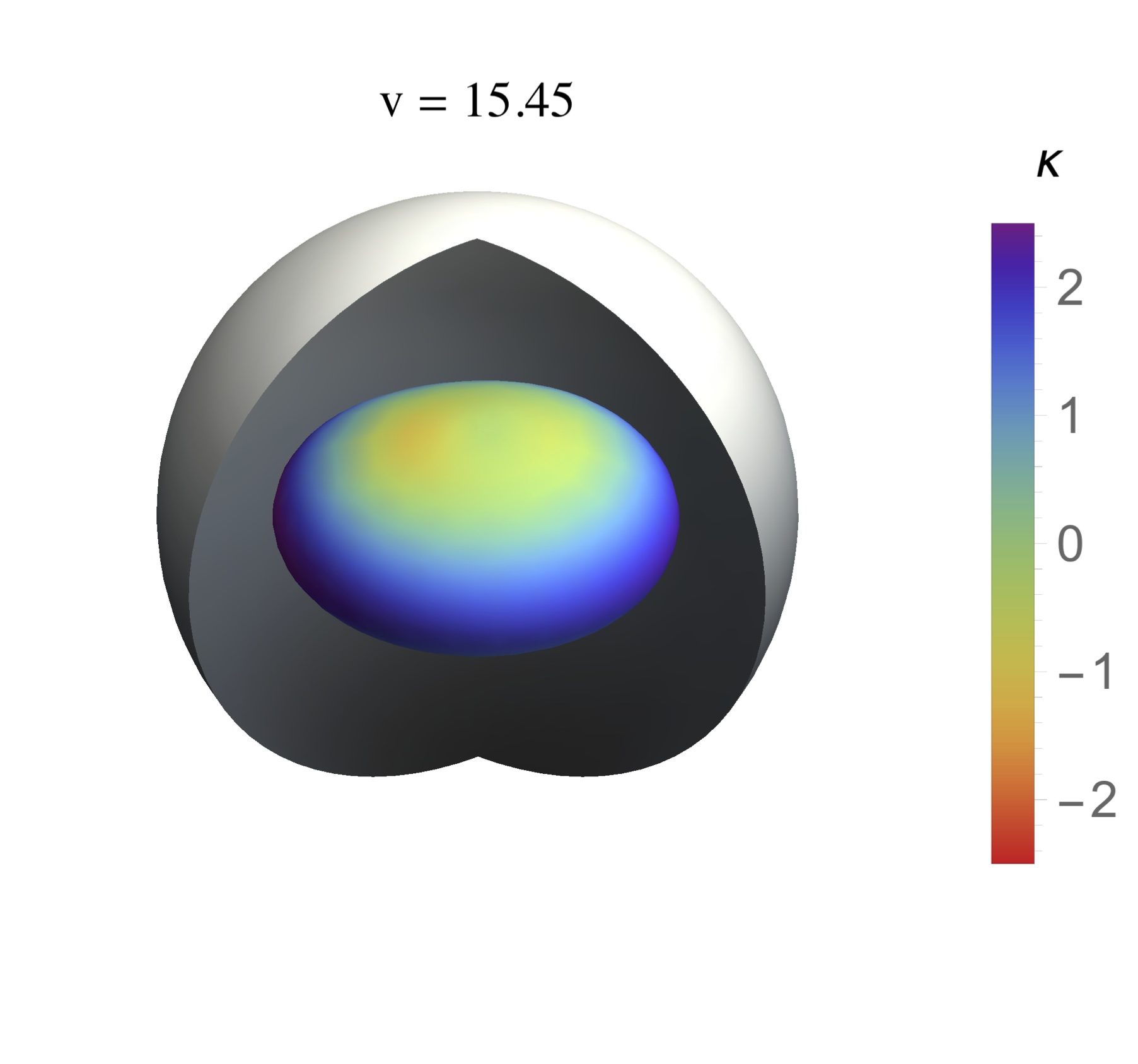}
\includegraphics[width=0.32\textwidth]{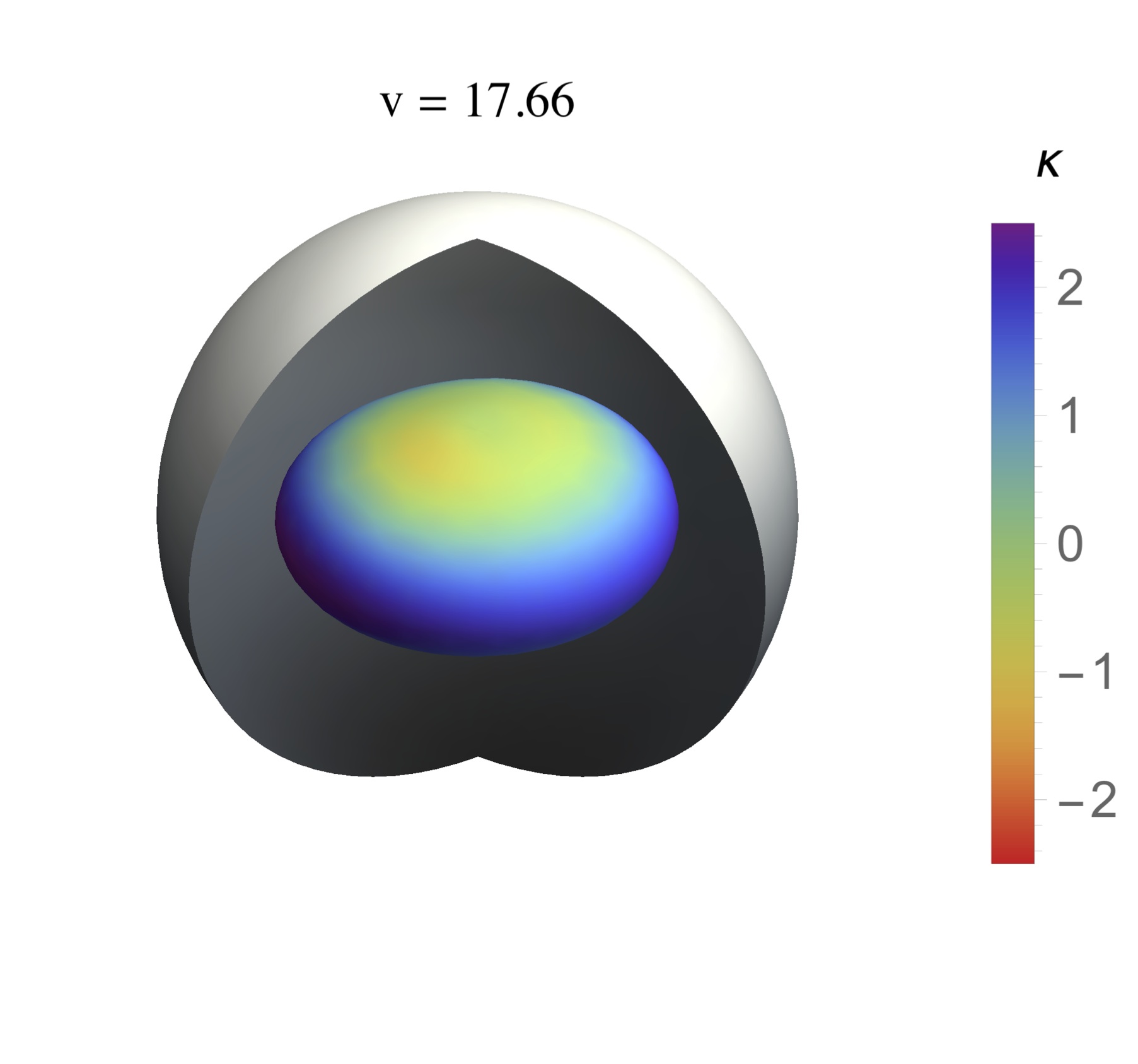}
\caption{Time series depicting the evolution of tidal bulges on an $a_*=0.9999$ black hole as its binary companion leaves periapsis.  The mass ratio is $\epsilon=1/125$ (see text for details).  The event horizon is embedded in the \poincare\ ball ($\ell=1$).  Colors indicate Gaussian curvature and the gray sphere is the hyperbolic boundary at $r=1$.  The black hole is negatively curved at the poles and it can also be negatively curved at the equator during periods of strong tidal distortion.}
\label{fig:bh9999}
\end{figure*}

Figure \ref{fig:bh9999} shows the horizon embedding we find at evenly spaced snapshots of constant ingoing time, $v$, as the perturbing body leaves periapsis.   The orbiting body raises tidal bulges on the black hole horizon.  Dissipation on the stretched horizon causes the black hole and the orbiting body to exchange energy, as discussed in detail in Refs.\ \cite{Hartle:1974gy,O'Sullivan:2014cba,O'Sullivan:2015lni}.  

Event horizons behave teleologically because their location depends on the entire future history of spacetime.  As a result, the shape of the horizon changes \emph{before} tidal forces are applied.  The timescale for teleological effects is inversely proportional to the surface gravity of the horizon.  The surface gravity is a decreasing function of spin, so teleological effects are most prominent at high spins.

\begin{figure}
\includegraphics[width=\columnwidth]{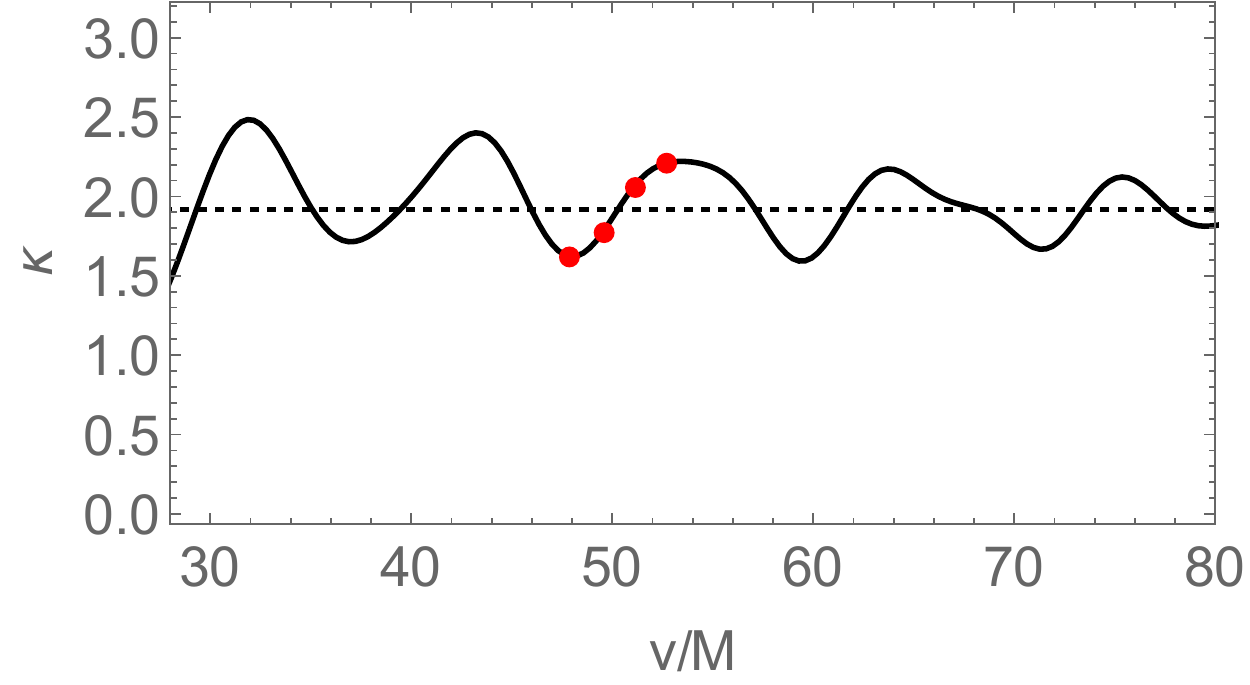}
\caption{Gaussian curvature of the $a_*=0.9999$ black hole at $\theta=\pi/2$ and $\phi=0$ as a function of ingoing time $v$.  The mass ratio has been increased to $\epsilon=1/50$.  The curvature of the unperturbed horizon is indicated (dotted line).  Red dots correspond to the embeddings shown in Figure \ref{fig:oscillate}.  The curvature oscillates despite the absence of significant tidal forces during this time interval, a phenomenon that was uncovered in Ref.\ \cite{O'Sullivan:2015lni}, but whose origin remains somewhat mysterious.}
\label{fig:datalist}
\end{figure}

\begin{figure*}
\includegraphics[width=0.23\textwidth]{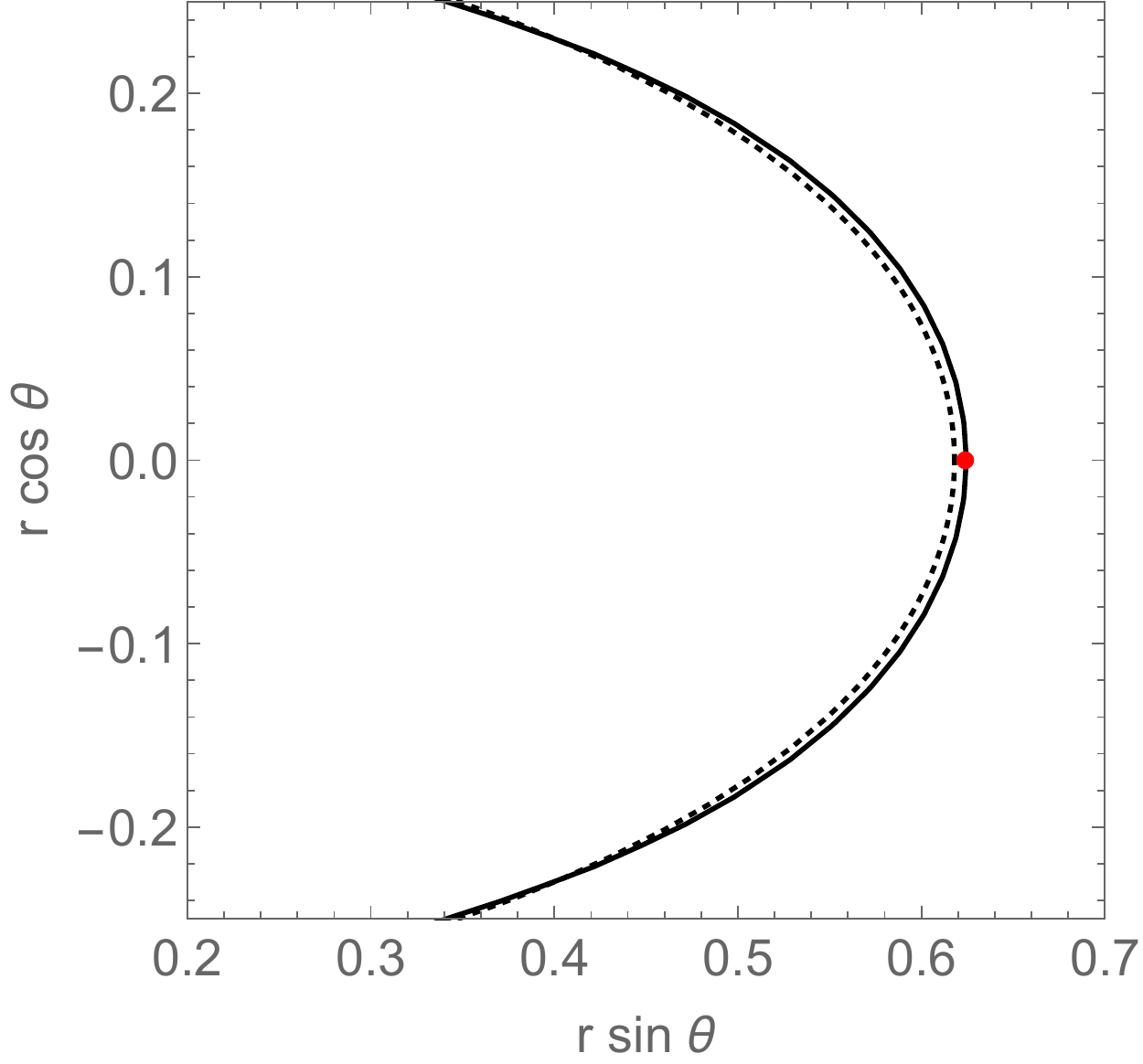}
\includegraphics[width=0.23\textwidth]{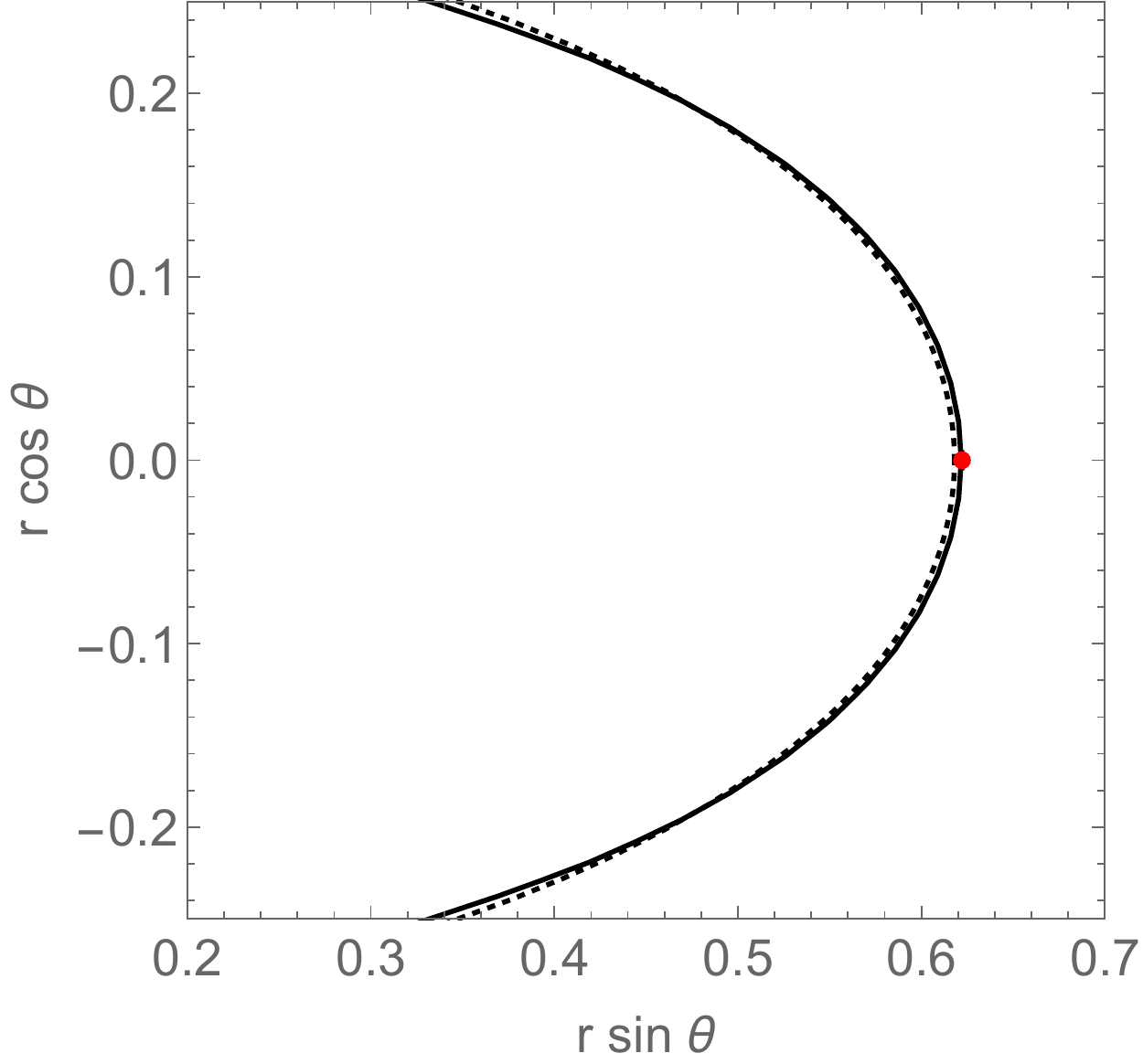}
\includegraphics[width=0.23\textwidth]{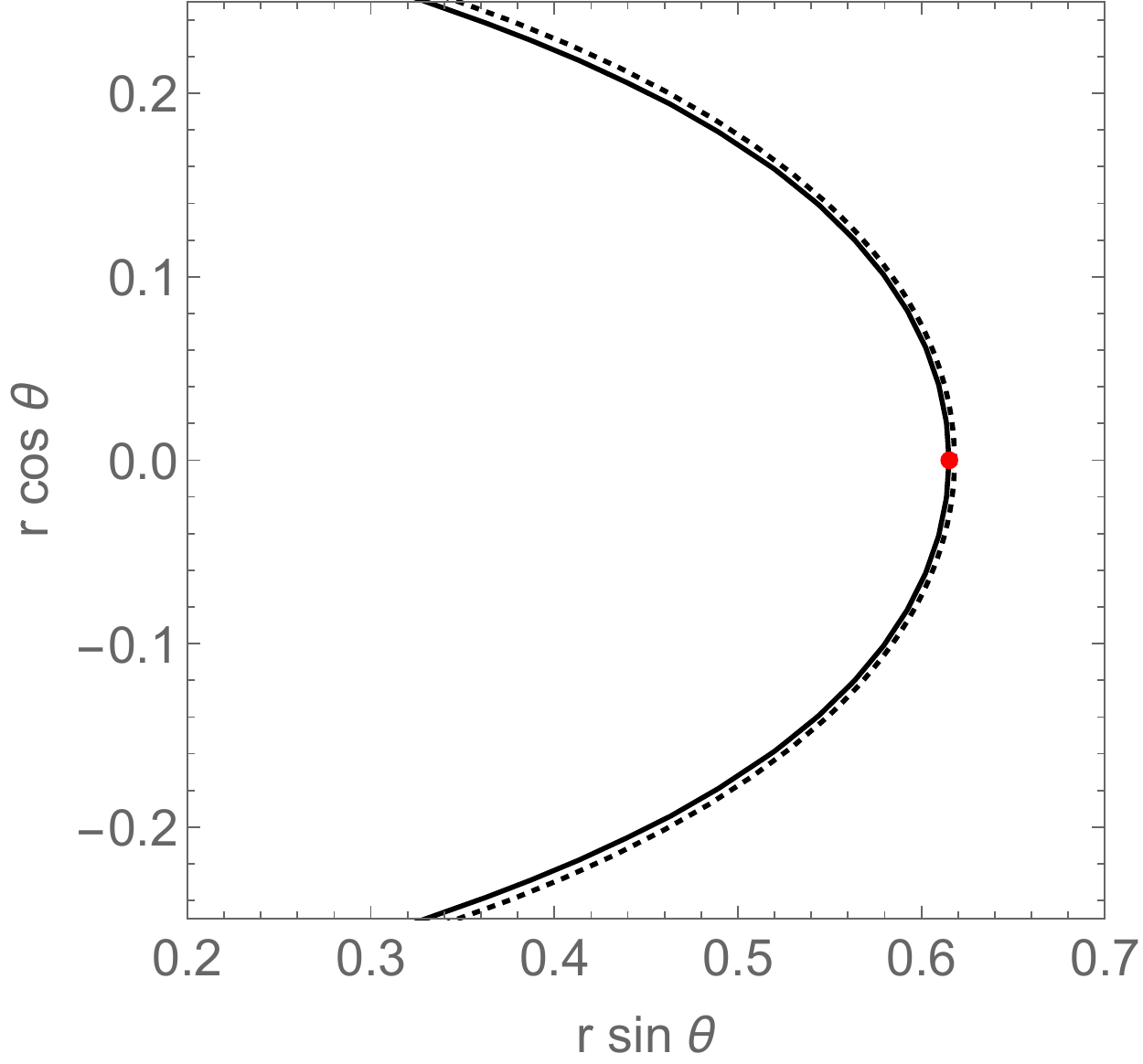}
\includegraphics[width=0.23\textwidth]{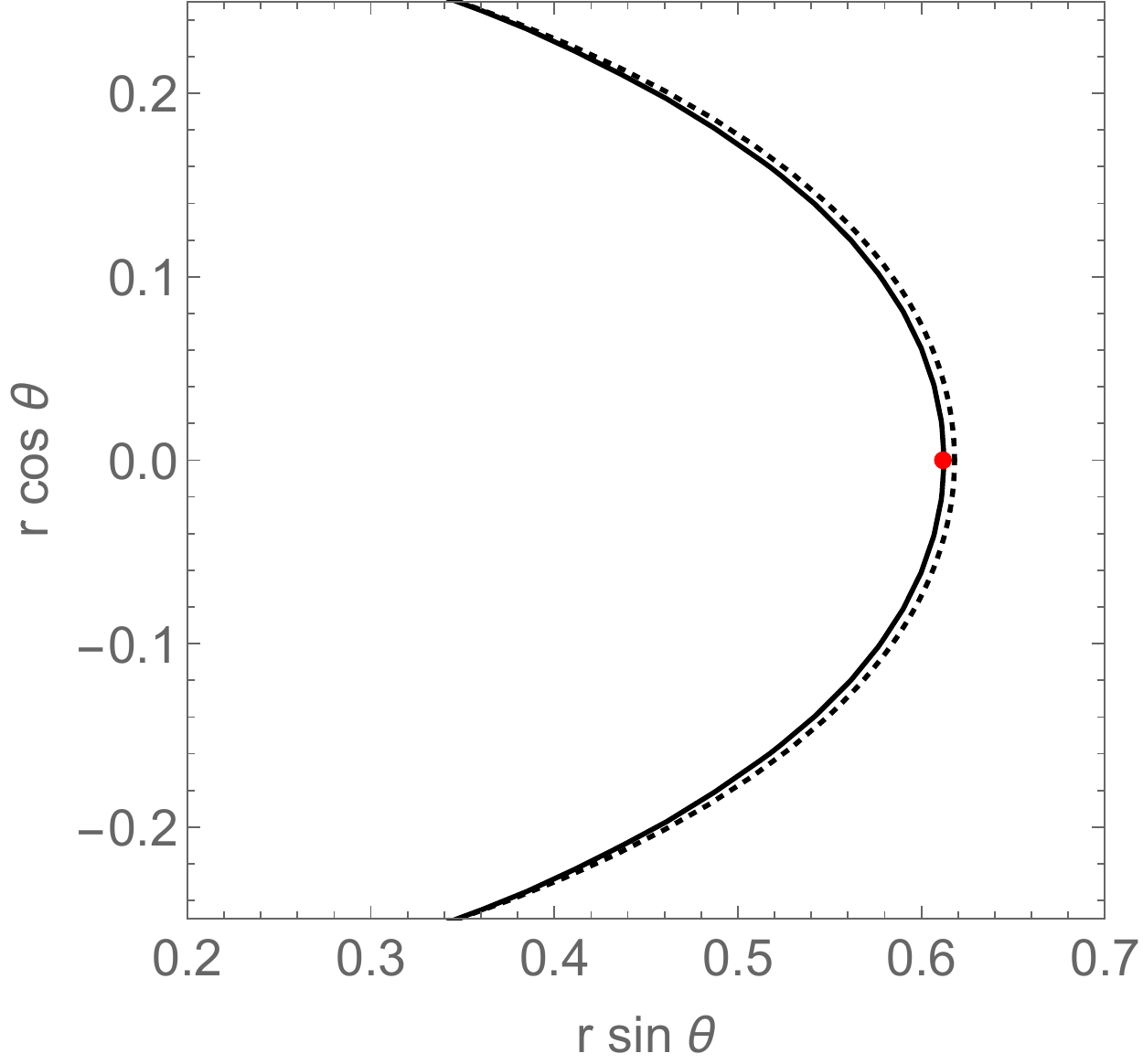}
\caption{Embeddings of the tidally distorted $a_*=0.9999$ horizon into \hthree\ (we have set $\phi=0$).  The perturbed (solid curves) and unperturbed (dotted curves) horizons are indicated.  Panels correspond to ingoing times $v/M=47.9, 49.5, 51.1$ and $52.7$ (left to right).   These times correspond to the red dots in Figure \ref{fig:datalist}.   The horizon oscillates about its unperturbed position despite the absence of significant tidal forces over this time interval.}
\label{fig:oscillate}
\end{figure*}

For the $a_*=0.9999$ case considered here, \cite{O'Sullivan:2015lni} observed that the horizon's curvature exhibits small amplitude, high-frequency oscillations.  During the oscillations themselves, there are no significant tidal forces acting on the horizon.  Figure \ref{fig:datalist} shows the Gaussian curvature of the horizon at $(\theta,\phi)=(\pi/2,0)$ as a function of ingoing time $v$, demonstrating the oscillations in Gaussian curvature shown in the earlier work.  (To make the effect more apparent, we have increased the mass ratio  to $\epsilon=1/50$).  Figure \ref{fig:oscillate} shows a visualizations of this effect on the horizon's geometry using embeddings into hyperbolic 3-space, a visualization which previous work could not provide.  We can see quite clearly that, whatever the physical origin of this process (which at this time remains somewhat mysterious), it can be clearly discerned in the horizon's embedded geometry.

\section{Concluding discussion}
\label{sec:conc}

In this paper, we have extended the tools we developed in previous work to allow us to visualize tidally distorted black hole horizons to encompass all allowed values of the Kerr spin parameter.  Our main innovation is to use a hyperbolic embedding space, as advocated in Ref.\ \cite{Gibbons:2009qe}, which circumvents the most critical shortcoming of a Euclidean embedding space (namely, that Euclidean spaces cannot be used to embed a region with negative curvature about an axis of rotational symmetry).  Using these tools, we are able to visualize the dynamics of horizon distortions for the particularly interesting case of black holes whose spins approach the maximal limit.  Past work has uncovered interesting and hard-to-understand behavior for these black holes.  The ability to visualize the horizon's behavior via these embeddings may be useful for understanding these phenomena.

Although we focus on embeddings of black holes tidally distorted by a small orbiting body in this analysis, we particularly wish to highlight the fact that hyperbolic spaces may be useful for a broader class of problems.  For example, these spaces may be useful for studies of rapidly rotating black holes in numerical relativity, or for studying the geometries of horizons following binary black hole coalescence.  Hyperbolic spaces are a natural arena for visualizing the behavior of surfaces with a wide range of Gaussian curvature, and so can provide powerful tools for presenting and understanding complex geometric phenomena, such as those involving black holes.

\begin{acknowledgments}

R.F.P.~was supported by a Prize Postdoctoral Fellowship in the Natural Sciences at Columbia University and a Pappalardo Fellowship in Physics at MIT.  S.A.H.~and S.O.~were supported by NSF grant PHY-1403261.  S.A.H.\ in addition thanks the Department of the School of Mathematics and Statistics of University College Dublin, where this paper was completed.

\end{acknowledgments}

\appendix

\section{Embeddings of unperturbed horizons}
\label{sec:embedapp}

It is helpful to rewrite the \poincare\ ball metric \eqref{eq:ball} in Cartesian coordinates,
\beq\label{eq:ballx}
ds_{{\mathbb{H}}^3}^2 =4\ell^2\frac{dx^2 + dy^2 + dz^2}{(1-r^2)^2}.
\eeq
Define a map $(\mu,\phi)\rightarrow (x,y,z)$ by the formulas
\beq
x=F(\mu)\cos\phi, \quad 
y=F(\mu)\sin\phi, \quad
z=G(\mu).
\eeq
Plugging into \eqref{eq:ballx} gives
\beq
ds^2 = 4\ell^2\frac{(F'^2 + G'^2)d\mu^2+F^2 d\phi^2}{(1-F^2 -G^2)^2}.
\eeq
Comparing this metric with the two-dimensional horizon metric \eqref{eq:gbar} gives
\begin{align}
4\ell^2 \frac{F(\mu)^2}{(1-F(\mu)^2 -G(\mu)^2)^2}&=\eta^2 f(\mu),\label{eq:feqn}\\
4\ell^2 \frac{F'(\mu)^2+G'(\mu)^2}{(1-F(\mu)^2 -G(\mu)^2)^2}&=\eta^2 f(\mu)^{-1}.\label{eq:geqn}
\end{align}
We solve \eqref{eq:feqn} for $F(\mu)$:
\beq
F = \sqrt{1+\frac{2\ell^2}{\eta^2f}-G^2
-\frac{2\sqrt{\ell^2(\ell^2+\eta^2f(1-G^2))}}{\eta^2f}}.
\eeq
The remaining equation is a nonlinear first order ordinary differential equation for $G(\mu)$, which we solve numerically.  
The embedding $(x,y,z) = (F(\mu)\cos\phi,F(\mu)\sin\phi,G(\mu))$ is now fully determined.

\section{Curvature formulas}
\label{sec:kw}

In this section we prove equations \eqref{eq:kw} and \eqref{eq:kwbar}.  The calculation  is in \cite{kazdan}, but we include it here for completeness.  
Let $h$ be a metric on $S^2$ with Gaussian curvature $\kappa$, and let $\hat{h} = e^{2v} h$ be a conformally related metric with Gaussian curvature $\hat{\kappa}$.  The claim is
\beq\label{eq:claim}
\hat{\kappa} = e^{-2v}(\kappa-\Delta v),
\eeq
where $\Delta$ is the Laplacian with respect to $h$.  

Eqs. \eqref{eq:kw} and \eqref{eq:kwbar} are special cases of \eqref{eq:claim}.   
The former is obtained by identifying $h=g_0$, $\hat{h}=(\bar{\zeta}^{-1})^*\bar{g}$, and $v=\bar{u}'$.  The latter is obtained by identifying $h=g_0$, $\hat{h}=(\zeta^{-1})^*g$, and $v=u'$.

To prove the claim, let $(\omega^1,\omega^2)$ be a local oriented orthonormal coframe field for $h$, and let $\phi^j_i=\Gamma^j_{ki}\omega^k$ be the connection form.  Cartan's equations of structure are
\begin{align}
d\omega^1 & = -\phi_{12} \wedge \omega^2,\label{eq:cartan1}\\
d\omega^2 & = \phi_{12} \wedge \omega^1,\label{eq:cartan2}\\
d\phi_{12} & = \kappa \omega^1 \wedge \omega^2\label{eq:cartan3}.
\end{align}

Now consider $(\hat{\omega}^1,\hat{\omega}^2)=(e^v\omega^1,e^v\omega^2)$, a local oriented orthonormal coframe field for $\hat{h}$.  Cartan's equation \eqref{eq:cartan1} gives
\begin{align}
d\hat{\omega}^1	&= e^v (dv \wedge \omega^1- \phi_{12}\wedge \omega^2)\notag\\
				&=({*dv}-\phi_{12})\wedge \hat{\omega}^2,
\end{align}
and it follows that $\hat{\phi}_{12} = \phi_{12} - {*dv}$.  So by Cartan's equation \eqref{eq:cartan3},
\begin{align}
\hat{\kappa} \hat{\omega}^1 \wedge \hat{\omega}^2 
	&= d\hat{\phi}_{12} = d\phi_{12} - d{*dv}\notag\\
	&=(\kappa-\Delta v) \omega^1\wedge \omega^2,
\end{align}
which implies the claim \eqref{eq:claim}.

\section{Distance formulas}
\label{sec:edge}

Equation \eqref{eq:dist1} defines distances between vertices using straight lines in \hthree.  We expect this to give the correct embedding in the large $N$ limit.  In practice  $N$ is finite, so it can be advantageous to improve the distance formula \eqref{eq:dist1} by, for instance, using geodesics on the embedded surface.  However, complicated distance formulas are computationally expensive. There is a trade-off between increasing the complexity of the distance formula \eqref{eq:dist1} and increasing $N$. 

Following \cite{nollert1996visualization}, we define distances between vertices using great circles on $S^2$.  Consider points on $S^2$ at polar angles $(\theta_p,\phi_p)$ and $(\theta_q,\phi_q)$.  They are connected by a great circle with arc length
\beq
\alpha = \cos^{-1}(\sin\theta_p\sin\theta_q\cos(\phi_p-\phi_q)+\cos{\theta_p}\cos{\theta_q}).
\eeq
The equation for the great circle  is 
\begin{align}
x_{pq} &= A \sin\theta_p\cos{\phi_p}+B \sin\theta_q\cos{\phi_q},\\
y_{pq} &= A \sin\theta_p\sin{\phi_p}+B \sin\theta_q\sin{\phi_q},\\
z_{pq} &= A \cos\theta_p+B \cos\theta_q,
\end{align}
where 
\beq
A= \frac{\sin((1-f)\alpha)}{\sin{\alpha}}, \quad B = \frac{\sin(f\alpha)}{\sin{\alpha}},
\eeq 
and $0\leq f \leq 1$ parametrizes distance along the circle.  The distance formula becomes
\beq\label{eq:dist2}
l^0_{pq} = \frac{4\ell^2}{(1-r^2)^2}
\left[
\left(\frac{\partial x_{pq}}{\partial f}\right)^2+\left(\frac{\partial y_{pq}}{\partial f}\right)^2+\left(\frac{\partial z_{pq}}{\partial f}\right)^2
\right],
\eeq
where derivatives are evaluated at the midpoint $f=0.5$.  For our numerical calculations, we used \eqref{eq:dist2} in place of \eqref{eq:dist1}.

\bibliography{ms}

\end{document}